\documentclass[]{spie}  

\usepackage{amsmath,amsfonts,amssymb}
\usepackage{graphicx}
\usepackage{float}
\usepackage[colorlinks=true, allcolors=blue]{hyperref}
\usepackage{siunitx}

\title{Optimized scintillation configuration for IMPISH hard x-ray detection}
\begin{document} 
\author[a]{Dorcas D. Oseni}
\author[a]{William S. Setterberg}
\author[a]{Reed B. Masek}
\author[a]{Lestat Clemmer}
\author[a]{Lindsay Glesener}
\author[a]{Philip Williams}
\author[b]{John Sample}
\author[c]{David M. Smith}
\author[d]{Amir Caspi}
\author[a]{Demoz Gebre-Egziabher}
\author[e]{Albert Shih}
\author[f]{Pascal Saint-Hilaire}
\affil[a]{University of Minnesota Twin Cities, Minneapolis, United States}
\affil[b]{Montana State University, Bozeman, United States}
\affil[c]{University of California, Santa Cruz, United States}
\affil[d]{Southwest Research Institute (SwRI), Boulder, United States}
\affil[e]{NASA Goddard Space Flight Center, Greenbelt, United States}
\affil[f]{University of California, Space Science Laboratory, Berkeley, United States}

\authorinfo{Further author information: \\D.D.O.: E-mail: oseni013@umn.edu, ORCID ID: https://orcid.org/0000-0003-4903-4626\\  L.G.: E-mail: glesener@umn.edu, ORCID ID: https://orcid.org/0000-0001-7092-2703\\
W.S.: E-mail: sette095@umn.edu, ORCID ID: https://orcid.org/0000-0003-2165-8314}

\pagestyle{empty} 
\setcounter{page}{301} 

\maketitle

\begin{abstract}

The Integrating Miniature Piggyback for Impulsive Solar Hard X-rays (IMPISH) is a solar X-ray spectrometer that features large-area scintillators, fast readout electronics, and good energy resolution in the hard X-ray band. IMPISH is a low-cost spectrometer designed to measure subsecond variation in hard X-ray time profiles from solar flares, with the goal of constraining particle acceleration timescales. To meet these requirements, we carried out a systematic optimization of the scintillation design, focusing on maximizing photon collection, reducing noise level, and improving energy resolution. We tested two high-yield scintillating crystals (LYSO, lutetium–yttrium oxyorthosilicate, and GAGG, gadolinium aluminum gallium garnet), two reflector types (specular and lambertian), two kinds of surface finishes (all sides polished and readout face only polished), different optical coupling materials and thicknesses, multiple geometries, and readout face angles. Our studies show that light collection improves with increasing the crystal's effective area, breaking its symmetry, and reducing the photon's travel length. These minimize photon loss due to self-absorption and total internal reflection. Through combined simulation and laboratory tests, we achieved an optimal energy resolution with the LYSO trapezoid-shaped crystal coupled to a Broadcom NUV-MT SiPM with a $\sim$0.25 mm thin Sylgard 184 optical pad.
\end{abstract}

\section{INTRODUCTION}
\label{sec:intro}  
\subsection{Science Goal}\label{sec:sscience_goal}

The Integrating Miniature Piggyback for Impulsive Solar Hard X-Rays (IMPISH) is a balloon piggyback mission
under the NASA Low Cost Access to Space (LCAS) program that will investigate electron acceleration in solar flares. It is a hard X-ray (HXR) spectrometer that will ideally fly for $>$1 week during a period of high solar activity to capture M- and X-class solar flares. IMPISH derives its heritage from the IMpulsive Phase Rapid Energetic Solar Spectrometer (IMPRESS)\cite{2022SPIE12181E..3MS, 2021PhDT.........2K} experiment that the University of Minnesota Twin Cities also develops. 

The IMPISH experiment is designed to investigate the mechanisms of electron acceleration through the quick bursts ($<$ 1 sec) present in solar flares\cite{2020ApJ...903...63K}. IMPISH will observe the spectra and timing of the spikes with high time resolution ($\sim$30 ms time-binned histograms), enabling us to derive the electron acceleration timescales, thus constraining the acceleration mechanisms. Fast measurement is necessary to capture the high rate of photons from solar flares, and the previous state-of-the-art solar-dedicated hard X-ray spacecraft, RHESSI, fell short of this quality due to its modulation of time signals at approximately 4 seconds. IMPISH is an instrument being developed as a collaboration between the University of Minnesota Twin Cities, 
Southwest Research Institute (SwRI), University of California, Santa Cruz (UCSC), and Montana State University.
Details about the solar flare science, the mission goals, and other components of IMPISH are further described in another 2025 SPIE paper\cite{reed2025the}.

\subsection{Hard X-ray Detection Mechanism}\label{sec:detection}

Hard X-ray detectors can be built from various materials, including semiconductor detectors---such as those used by RHESSI\cite{smith2002rhessi}, GRIPS\cite{shih2012gamma} and Solar Orbiter/STIX\cite{krucker2020spectrometer}---and scintillating detectors, like those used by Fermi/GBM\cite{meegan2009fermi}, IMPRESS\cite{2022SPIE12181E..3MS}  and IMPISH\cite{reed2025the}. 

Semiconductor detectors generally have better energy resolution, but they are limited in their photo-absorption capabilities at high energies and in their collecting area. We chose scintillators because they are better suited for detecting hard X-rays with a large effective area, which is needed due to significant attenuation of hard X-rays at balloon altitudes, and the faintness of high-energy flare spectra. Moreover, our primary goal of measuring non-thermal power laws does not require the fine energy resolution offered by semiconductor detectors.

In scintillation detectors, the process starts with the crystals converting X-rays to optical or UV photons through photoelectric absorption. The number of scintillated photons is proportional to the X-ray energy, and they are detected using photomultiplier tubes or photodiodes. For IMPISH, the photodiode pixels used are small avalanche photodiode (APD) cells made from silicon substrate, hence referred to as Silicon Photomultipliers (SiPMs).

This paper presents our work on optimizing the performance of scintillating detectors as a function of crystal types, geometry, surface finish, reflectors, and readout face angle. 
This work prioritizes the development of compact detectors featuring large scintillator areas with relatively few readout channels, fast timing resolution, sensitivity to a broad range of X-rays (10–100 keV), and good energy resolution in that range. 

\section{IMPISH DETECTOR DESIGN CONSIDERATIONS}\label{IMPISH}
IMPISH is a scintillator-based instrument that uses inorganic scintillators doped with Cerium. 

\subsection{Crystal}

The crystal to be selected for hard X-ray detection is guided by several important requirements: high light yield, being non-hygroscopic for easy handling in the lab, fast rise and decay times for high rate detection, and ultimately good energy resolution. Two prominent candidates that satisfy most of these criteria are cerium-doped lutetium yttrium orthosilicate (LYSO:Ce) and the newer gadolinium aluminium gallium garnet (GAGG:Ce). The crystals used in the tests reported here were supplied by Epic Crystal Co.

GAGG has a higher light yield of approximately 54 photons/keV, which roughly doubles LYSO's yield of 30 photons/keV. However, the LYSO emission peak at approximately 420 nm perfectly matches the sensitivity peak of the SiPMs we used, at approximately 420 nm. Contrast this with the GAGG emission peak at 540 nm, where the SiPM is half as sensitive. This mismatch in spectral response limits the light collection efficiency of GAGG in this setup, despite its higher light yield. To maximize the performance of GAGG, alternate photomultipliers would need to be used.

GAGG has an afterglow effect due to impurities or defects within the crystal lattice. These impurity sites temporarily prevent excited carriers (electrons or holes) from transitioning to the ground state.
The electron eventually recombines on luminescence sites, causing delayed luminescence. The resulting slow component is referred to as afterglow\cite{2000rdm..book.....K, 2011LanB.21B1...45L}. The slow decay can add up and result in pileup since we are observing a high-rate source (the Sun), and the afterglow can mimic or obscure low-energy events in the detector.

Considering their physical characteristics, LYSO has some advantages. GAGG has a density of 6.6 g/cm$^3$ and an effective atomic number of 54.4 (Table \ref{tab:gagglysoyap}). LYSO has a higher density of 7.25 g/cm$^3$ and a higher effective atomic density of 66, both of which are advantageous for X-ray detection due to increased stopping power\cite{2004ITNS...51.1084P, 2013ITNS...60..988Y}. Additionally, LYSO exhibits a faster decay time of $\sim$40 ns compared to GAGG's decay time of $\sim$90 ns, which helps reduce pileup events, which is critical for spectral measurements of bright solar flares. Although pileup is an unlikely issue for the IMPISH mission, due to atmospheric attenuation at balloon altitudes, a future spacecraft version of this instrument would benefit from this fast detection.

Yttrium Aluminum Perovskite (Cerium)—YAP(Ce)—is another strong contender with a faster decay time of 25 ns and a moderate light yield of $\sim$18 photons/keV. However, its emission peak of 370 nm lies outside the optimal sensitivity range of the Onsemi C-series SiPM we used for LYSO and GAGG. Instead, we tested a YAP crystal with a Broadcom near-ultraviolet NUV-MT silicon photomultiplier (SiPM), which has higher sensitivity in the near-UV range down to $\sim$250 nm. These tests are ongoing and will be shown in future work.

\begin{table}[H]
\caption{The physical properties of LYSO, GAGG, and YAP compared to CeBr used by the IMPRESS cubesat and NaI, a commonly used scintillating crystal for astrophysical purposes.} 
\label{tab:gagglysoyap}
\begin{center}       
\begin{tabular}{llllll} 
\hline
\rule[-1ex]{0pt}{3.5ex}  Scintillator & LYSO & GAGG & YAP & CeBr$_3$ & NaI \\
\hline
\rule[-1ex]{0pt}{3.5ex}  Density [g/cm$^3$] & 7.25  & 6.60 & 5.40 & 5.10 & 3.67\\
\rule[-1ex]{0pt}{3.5ex}  Light yield [photons/keV] & 30 & 54  &18 & 60 & 38\\
\rule[-1ex]{0pt}{3.5ex}  Decay constant [ns] & 40 & 90 & 25 & 20 & 264\\
\rule[-1ex]{0pt}{3.5ex}  Peak emission [nm] & 420 & 530  & 370 & 380 & 415 \\
\rule[-1ex]{0pt}{3.5ex}  Effective atomic number & 54.4 & 66 & 36 & 45.9 & 56\\
\hline
\end{tabular}
\end{center}
\end{table}

\subsection{Photodetection and Readout Electronics}

The silicon photomultiplier (SiPM) was selected as our photodetector due to its electronic advantages of mechanical compactness, high internal gain ($\sim$10$^{6}$), insensitivity to magnetic fields, ability to operate at low bias voltage, low noise, high photon detection efficiency, and easy soldering. While Onsemi C-series 6 mm × 6 mm SiPMs were used for initial testing, the flight model will contain Broadcom’s NUV-MT AFBR-S4N66P014M SiPMs, which are optimized for near-UV sensitivity. The Broadcom NUV-MT SiPM was ultimately chosen for the flight detector because of its higher peak photon detection efficiency (PDE) of about 63\% compared to Onsemi's 41\%. This makes it a better match not only for YAP but also for maximizing performance with LYSO under solar hard X-ray conditions. 

SiPM is a large array of Geiger-mode avalance photodiodes (APDs). A reverse bias voltage above the breakdown voltage is applied across the APDs, resulting in an avalanche multiplication of electrons that creates a current output. The SiPM output current is amplified and converted to voltage using an Analog Devices LTC6268-10 preamplifier in a transimpedance configuration and buffered by a Texas Instruments BUF602. The signals were processed in real-time by the off-the-shelf Bridgeport Instruments SiPM-3000 module, which includes a 40 MSPS ADC and an AMD/Xilinx Spartan-6 FPGA. The system integrates the digitized pulse over time, and this integral is proportional to the total charge delivered by the SiPM, and thus to the energy deposited in the scintillator. The resulting histogram of integrated pulse values forms an energy spectrum, where the ADC bin position reflects the combination of the original X-ray energy, scintillator light yield, SiPM photon detection efficiency, and electronic settings such as bias voltage, integration time, and digital gain.

Some of the properties we prioritized in the scintillating detector are high scintillator light yield, fast scintillation decay time, effective optical coupling between the scintillator and SiPM (compatible refraction indices) to reduce total internal reflection at the readout face, well-matched scintillator emission wavelength to SiPM sensitivity, reduced self-absorption within the crystal, linearity of crystal light yield, reduced or absent afterglow/phosphorescence of scintillators, and good SiPM quantum efficiency.

These parameters are interdependent, and this work is focused on optimizing them through laboratory testing and simulations, starting with the crystal types. The other components of the IMPISH mission, beyond the scintillating detector, are discussed in our second 2025 SPIE paper\cite{reed2025the}.

\subsection{X-ray Sources}

To evaluate the performance of the IMPISH detector, two standard radioactive sources were used: $^{133}$Ba and $^{241}$Am. 
The $^{133}$Ba emission line at 81 keV and $^{241}$Am emission line at 59.5 keV are used to characterize the IMPISH detector.

\subsection{Optical Guidance}

To guide the maximum number of optical photons from the scintillator to the silicon photomultiplier, the following parameters were optimized. 

\subsubsection{Surface Finish}

The efficiency of photon arrival at the photodetector is influenced by reflection probability within the crystal, which is controlled by surface roughness.  The surface finish affects both the number of photon reflections and the angular distribution of those reflections, which in turn influences the likelihood of photon escape or reabsorption. 

Two surface finishes were tested: all sides polished (ASP), where all six faces of the crystal are optically polished, and one side polished (1SP), where only the readout face (coupled with the SiPM) is polished; the remaining five faces are left unpolished.

Previous studies with an LSO\cite{2013PMB....58.2185R} (Lutetium Oxyorthosilicate) crystal have shown that surface roughness affects angular reflection characteristics. Rough surfaces tend to increase reflection at near-normal incidence angles but reduce it at higher angles. In contrast, mechanically polished surfaces behave like a flat surface and allow total internal reflection at incident angles greater than the critical angle, which is approximately 33$^{\circ}$ for a LYSO-, LSO- and YAP-air surface. 
Also, a comparison of angular reflectance at a fixed incidence angle (e.g., 45$^{\circ}$), done in the paper\cite{2013PMB....58.2185R}, shows that the rough surface produces dispersed rays while the polished surface yields ideal specular reflection. The geometry and absorption property of the crystal determine whether diffuse or specular reflection results in more photon loss.

Hence, optimizing the surface finish is a trade-off between maximizing light output at the SiPM face and minimizing photon loss to scattering and reabsorption.

\subsubsection{Geometry}

\begin{figure} [ht]
   \begin{center}
   \begin{tabular}{c} 
   \includegraphics[width=10cm]{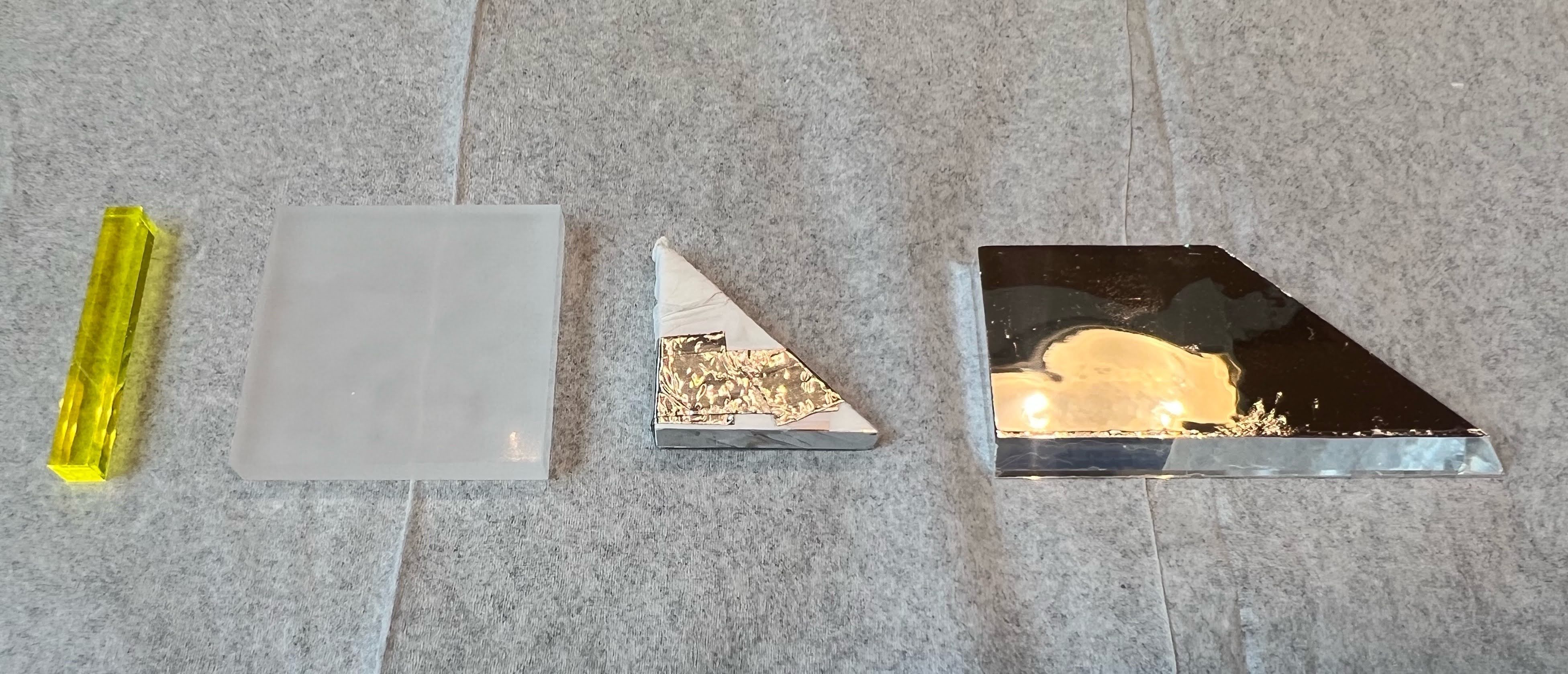}
   \end{tabular}
   \end{center}
   \caption[example]
   { \label{fig:shape}
   \textit{Left} to \textit{right}: GAGG rod all-side polished, LYSO plate one-side polished, triangle clad in enhanced specular reflector (ESR) and then tightly wrapped with teflon tape, trapezoid crystal with 45$^{\circ}$ angled readout face clad in ESR.}
   \end{figure}

Crystal geometry is one of the least explored constraints for maximizing the light collection efficiency of scintillating detectors. In this study, various geometries and different readout face angles were tested, some of which are shown in Figure \ref{fig:shape}. The tested geometries included rods (5 mm $\times$ 5 mm $\times$ 40 mm), plates (40 mm $\times$ 5 mm $\times$ 40 mm), triangles (45 mm $\times$ 5 mm $\times$ 40 mm), and trapezoids (66 mm $\times$ 5 mm $\times$ 33 mm).

Previous studies have shown that energy resolution improves with increasing aspect ratio (the ratio of the length of the readout face to its height) of the scintillating crystal, and that breaking the symmetry improves light yield. A previous study\cite{1999NIMPA.437..374H} that tested maximum light output of narrow LSO crystals as a function of geometry and surface finish found a weak dependence on crystal length and surface finish, and another\cite{2012ITNS...59.2340P} showed that thin rods have lower light yield compared to bigger crystals.

Additionally, we examined readout face angles, comparing flat and angled surfaces. This is to minimize total internal reflections of photons at the readout face and increase readout face area.

\subsubsection{Reflectors}
Because the scintillation light is emitted isotropically from the X-ray interaction site, only a limited fraction of scintillated photons will travel directly to the SiPM. Two different kinds of reflectors were tested to guide the optical photons to the readout face: an Enhanced Specular Reflector (ESR) with specular reflectivity and Teflon with Lambertian reflectivity. For a specular reflector, the angle of incidence is equal to the angle of reflection $(\theta_i = \theta_r)$, while light making contact with Lambertian surfaces reflect light diffusely in a pattern proportional to the cosine of the angle between the direction of observation and the surface normal (I($\theta$) $\propto$ $\cos$($\theta$)). The two reflectors have $>$98\% reflection across the visible spectrum, and they help guide optical photons to the SiPM.

Photon collection at the readout face is controlled by the properties of total internal reflection.
\begin{itemize}
    \item If the incident angle = critical angle, there is refraction along the boundary, potentially causing photon escape.
    \item If the incident angle $>$ critical angle, there is total internal reflection. We would want this to happen at the sides of the crystal, but not at the readout face.
    \item If the incident angle $<$ critical angle, partial transmission and partial reflection occur. The reflectors acting as an external mirror would be useful in recovering escaping photons in this case.
\end{itemize}

The two reflector types were tested to determine which would yield a better energy resolution performance in each of our crystal geometries.

\subsubsection{Optical Coupling}
Work was done to ensure an optimal setup for the silicone optical pads that couple the crystal readout faces to the SiPMs. 
It was recommended that the thickness of these pads be minimized to increase light collection at the SiPMs. Optical pads are commonly sold in thicknesses down to only 1 mm, so a method to create thinner optical pads was developed. Prior work suggested using Sylgard 184. A custom optical pad with a thickness of $\sim$0.25 mm was created and tested against an Eljen Technology, 1 mm thick, EJ-560 optical pad.

\subsubsection{Photon Loss by Self-Absorption}

Self-absorption occurs when photons generated within the scintillator are reabsorbed before reaching the photodetector, reducing the number of detected photons and degrading detector performance. While self-absorption is typically negligible in small crystals, it becomes a significant loss mechanism in larger crystals spanning several centimeters \cite{2000rdm..book.....K}. This loss directly impacts the light yield and energy resolution, making it critical to understand and minimize it for optimal detector efficiency.

This factor couples strongly with other factors of the crystal---especially geometry and reflective qualities---to influence light collection.  For example, diffuse reflection can be utilized to ensure that scintillated photons do not enter a "reflective trap" in which they are continually reflected back and forth across a large dimension of the crystal until they are absorbed in the crystal.

\subsection{Energy Resolution}
The number of detected scintillated photons from the same X-ray energy varies from event to event due to statistical fluctuations. While this follows Poisson statistics\cite{2011LanB.21B1...45L}, the large photon count allows it to be approximated by a Gaussian distribution.

Energy resolution quantifies how well a scintillator can distinguish between different energies and is derived from the Gaussian fit of the energy peak. It is defined as the difference between the two x-values at which the function reaches half of its maximum value; hence, it is called the full width at half maximum (FWHM). The energy resolution values in this paper are calculated using
\begin{equation}\label{eq:res}
    \text{Relative resolution} =\frac{\text{FWHM}}{\mu} \times 100
\end{equation}
expressed in percentage, where $\mu$ is x-value of the mean (i.e. the position of the Gaussian peak). Lower resolution percentages indicate sharper peaks and better energy discrimination.

\subsection{Intrinsic Resolution and Non-Proportional Light Yield}

The best possible energy resolution achievable by a scintillating crystal under ideal conditions is independent of the photodetector and the electronic components but controlled by the \textit{intrinsic resolution} of the crystal. It is an umbrella term that captures the non-linear properties of scintillating crystals and other fundamental properties. A previous study\cite{2009ITNS...56.3800C} shows us that LYSO is greatly affected by its non-linear characteristic when compared to other crystals. Its energy resolution broadens by about a factor of five beyond pure Poisson statistics due to its non-proportional response. In contrast, YAP:Ce exhibits only about a twofold broadening, offering superior intrinsic resolution, especially in our energy range of interest ($\sim$59.5 keV) \cite{gektin2017inorganic, 2017isds.book.....L}. This motivates our future consideration of YAP:Ce to enhance the overall energy resolution of scintillating detectors, and that work will be reported in a future paper.

Non-proportionality in light yield is also a limitation to the crystal's energy resolution and is greatly influenced by the properties of the host crystal and the doping agents\cite{chewpraditkul2009scintillation, 2009ITNS...56.2499S}. It can be due to Compton scattering\cite{1961NucIM..11..340I}, which is beyond our energy range of interest, as well as absorption at the characteristic lines of the scintillator materials\cite{1995RadM...24..355D}.

\subsection{Dark Count Rate}
In the absence of illumination, the SiPM generates thermal electrons termed dark counts. They are indistinguishable from actual photon counts and are significantly influenced by temperature and overvoltage. Photodiodes operating well above room temperature will experience a rapid rise in dark current, which corresponds to the sum of dark counts generated. This is not an issue for missions that operate at temperatures lower than or comparable to room temperature. 

In addition, increasing the number of SiPMs in a channel results in higher dark count rates. The rod configuration tested requires only one SiPM, but the plate and triangle configurations require six, and the trapezoid configuration requires ten SiPMs. As a result, a trade-off must be made between light collection efficiency and the effective area provided by the explored scintillator geometries and their associated SiPM dark count rates.

\subsection{Simulations: \texttt{pvtrace} and Geant4}\label{sec:simulation}

We utilized two simulation tools to better understand photon transport within the detector system and optimize geometry and surface properties. Geant4 (Geometry and Tracking)\cite{2003NIMPA.506..250A, 2006ITNS...53..270A, 2016NIMPA.835..186A} is a widely used Monte Carlo framework for simulating the movement of particles through matter, and is heavily used in different areas of physics, especially by the high-energy particle physics community. We specifically applied it to simulate the behaviors of optical photon within the crystal. By setting parameters like the surface finish, reflectors, crystal photon yield, etc, we can simulate the optical photon interactions and estimate the light collection efficiency of the system.

We also used a Python package called \verb|pvtrace|, developed by Daniel Farell\cite{farrell2008characterising}, to ray trace photons in the scintillator and estimate how many photons reach the SiPM from varying crystal shapes. By understanding that self-absorption by the crystal and total internal reflection at the readout face play a large role in the signal produced, we simulated as many shapes as possible using \verb|pvtrace|.
The configurations were modeled to recreate the setup used in the lab. We have a 1 mm thick optical pad between the crystal and the SiPM.
The ESR shells are constructed such that there is a 1 $\mu$m gap between the crystal and the ESR, and the ESR is 0.1 mm thick. The refractive indices used were 1.8 for the LYSO crystal, 1.4 for the optical pad, and 1.6 for the SiPM enclosure. An absorption coefficient of  \SI{50}{\per\centi\meter}, as reported in a previous study\cite{2006NIMPA.564..506V}, was used for the LYSO simulations in Geant4 and \texttt{pvtrace}.

It is important to know that these simulation tools were utilized under different modeling assumptions. \verb|pvtrace| assumes 100\% reflectivity (although there is a 5\% chance of Lambertian reflection), thus primarily focusing on geometrical optics. In contrast, Geant4 allows incorporating more processes (e.g., scintillation processes and detailed material properties). For Geant4, we conducted two types of simulations: one including full scintillation physics (modeling ionizing events and the resulting light emission) and another with direct firing of optical photons (no ionizing events).

\section{RESULTS}

\subsection{LYSO vs GAGG}

\begin{figure} [ht]
   \begin{center}
   \begin{tabular}{c} 
   \includegraphics[height=5cm]{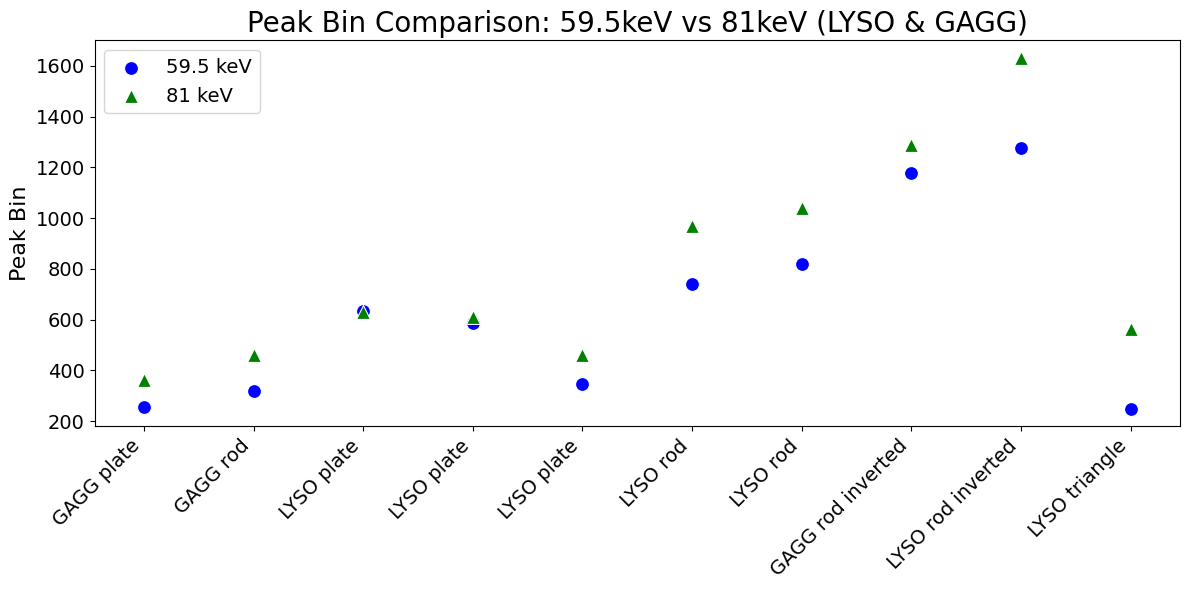}\\
   \includegraphics[height=5cm]{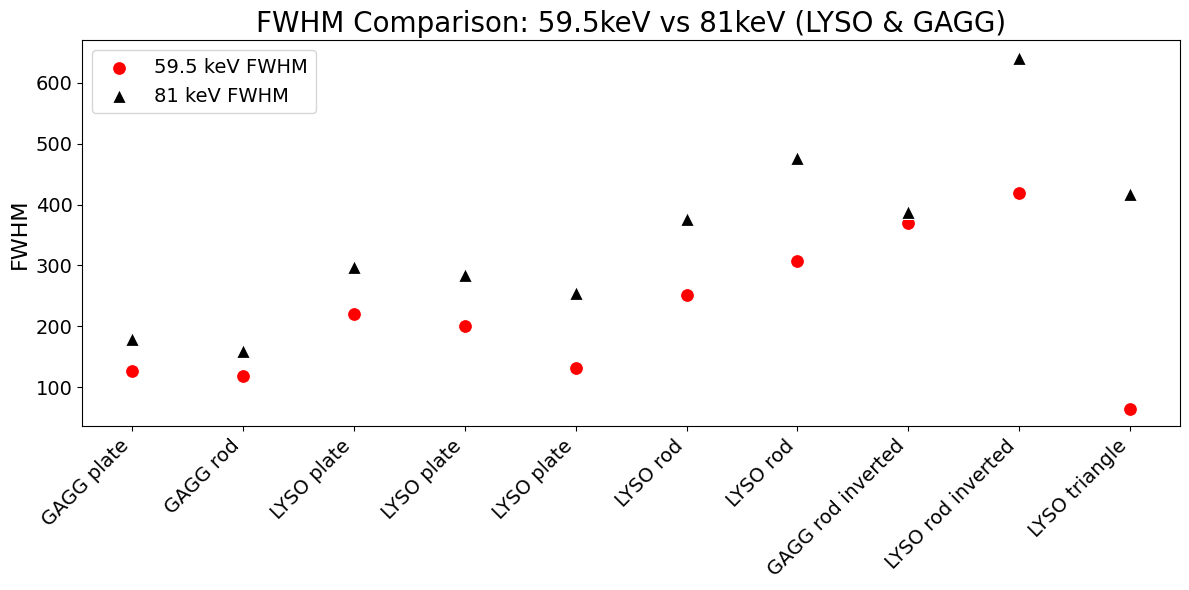}\\
   \includegraphics[height=5cm]{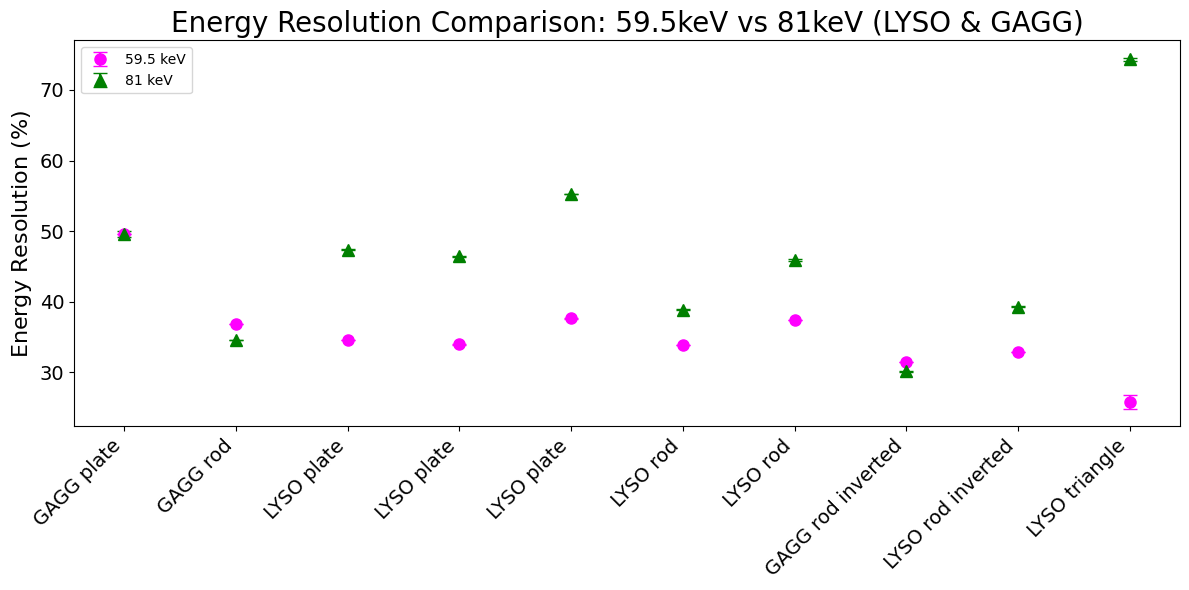}
   \end{tabular}
   \end{center}
   \caption[example]
   { \label{fig:lysoGAGG}
   \textit{Top to bottom:} Peak ADC bins, full width at half maximum (FWHM), and energy resolution with error bars (statistical fit error) are shown for different configurations of the scintillating detector. The triangle marker represents the 81 keV line ($^{133}$Ba) and the circle marker is the 59.5 keV line ($^{241}$Am). The "repeated" LYSO configurations corresponds to crystals of the same type and shape but with different surface finishes or reflectors. Some configurations with higher peak bin values were offset by higher FWHM, resulting in comparable resolution, especially for 59.5 keV peaks.}
   \end{figure}

Scintillated light collection influences the peak position on the ADC bin axis, while uniformity in light collection, alongside any other source of energy resolution, influences the width of pulse distributions. We tested multiple configurations like LYSO ASP plate, LYSO 1SP plate, LYSO ASP rod, LYSO 1SP rod, LYSO ASP triangle, LYSO ASP trapezoid, GAGG ASP plate, GAGG ASP rod, GAGG 1SP rod; and each of these were either wrapped with Teflon or ESR reflector. Figure \ref{fig:lysoGAGG} (top panel) compares the distribution of the location of the energy peak on the ADC bin for LYSO and GAGG configurations when exposed to  $^{133}$Ba (81 keV) and $^{241}$Am (59.5 keV) sources, under identical detector settings. The "repeated" LYSO configurations are with different surface finish or reflectors, but are not shown here to put focus on LYSO vs GAGG values. LYSO shows the 59.5 and 81 keV peaks at higher ADC bins compared to GAGG, despite the fact that GAGG has nearly double the light yield. This result is likely due to the better match between LYSO spectra and SiPM spectral sensitivity.

However, LYSO also shows broader energy peak widths (Figure \ref{fig:lysoGAGG}, middle panel), which could be due to reduced uniformity in light collection. As a result, the energy resolution, defined in Equation \ref{eq:res}, is comparable between LYSO and GAGG at 59.5 keV (Figure \ref{fig:lysoGAGG}, bottom panel), with most values between 30 and 40\%.

Interestingly, LYSO shows a consistently better resolution at lower energy, 59.5 keV ($^{241}$Am), than at higher energy, 81 keV ($^{133}$Ba), which is counterintuitive. In general, energy resolution should improve at higher energies due to an increased number of scintillation photons, if the resolution is largely governed by the counting statistics of the collected light. This anomaly may be due to the non-proportional light yield property of LYSO at low energies\cite{2013NIMPA.712...34I, wanarak2012light}, or may indicate that other resolution components have larger contributions.

Moreover, GAGG suffers from long-lived afterglow emission, which adds low-level noise to the baseline and introduces trailing signal components, which degrade performance in fast timing applications like IMPISH, especially at low energy. Due to its better timing response and spectral match with the SiPM, LYSO was ultimately selected as the preferred scintillator for our detector.

\subsection{Roughness}

To evaluate the effect of surface finish on the detector performance, we compared the energy resolution for LYSO crystals with one side polished (1SP) and all sides polished (ASP) cases. In both cases, the crystals were wrapped with Teflon reflectors and measured under similar conditions using $^{241}$Am and $^{133}$Ba sources. As shown in Table \ref{tab:ER_rough} (and visualized in Figure \ref{fig:rough} in the Appendix), the energy resolutions were almost the same. For the 59.5 and 81 keV peaks, the ASP crystal has a slightly better resolution compared to the 1SP crystal. Interestingly, this result contrasts with earlier findings such as Ishibashi et al. (1986)\cite{1986JaJAP..25.1435I}, which suggested that roughened surfaces can enhance light collection uniformity. The uncertainty in the resolution, shown in Table \ref{tab:ER_rough}, was calculated using error propagation, including the variances of the fitted Gaussian parameters, peak position ($\mu$) and width ($\sigma$), as well as their covariance.

\begin{table}[H]
\caption{Energy resolution for LYSO crystals wrapped with Teflon reflectors, comparing crystals with all sides polished (ASP) and one side polished (1SP). The uncertainties reflect statistical errors from Gaussian fits only and do not include potential systematic contributions.} 
\label{tab:ER_rough}
\begin{center}       
\begin{tabular}{lll} 
\hline
\rule[-1ex]{0pt}{3.5ex}  Polish/Energy Resolution (\%) & $^{241}$Am (59.5 keV) & $^{133}$Ba (81 keV) \\
\hline
\rule[-1ex]{0pt}{3.5ex}  One side polished & 34.63 ± 0.15  & 47.39 ± 0.24   \\
\rule[-1ex]{0pt}{3.5ex}  All sides polished & 33.99 ± 0.10 & 46.42 ± 0.21  \\
\hline
\end{tabular}
\end{center}
\end{table}

\subsection{Reflectors}

Early testing revealed that achieving an airtight and uniform connection between the reflectors and crystals was essential for greatly reducing photon loss. Particularly, energy resolution and light collection improved significantly when crystals wrapped with ESR film were further tightly sealed with Teflon tape, compared to those with ESR cladding alone. 
We used the LYSO ASP plate to compare the performance of Teflon (Lambertian reflectivity) and ESR (specular reflectivity) under similar testing conditions.

The Teflon-clad crystal performed better than the ESR-wrapped configuration for both energy peaks tested. For the 81 keV peak ($^{133}$Ba), Teflon improved energy resolution by approximately 10\%. At 59.5 keV ($^{241}$Am), the Teflon-clad crystal showed a 19\% improvement (see Figure \ref{fig:reflect} in the Appendix).

\subsection{Geometry}

\subsubsection{Shape}
The tests compared the square plates and rod configurations, with the rod read out on the shorter side (5 $\times$ 5 mm). The rods have better energy resolution (i.e. a narrower energy peak), which is attributed to more uniform light collection. However, fewer scintillated photons reach the SiPM, as expected, due to the smaller active area. Despite the rod having more consistent and better energy resolution, plates provide a larger X-ray collection area, which is essential since hard X-ray flux is fainter at our point of observation due to the Earth’s atmosphere. Additionally, the plates are easier to mount to the SiPMs since an  arrangement of multiple rods is not ideal from a mechanical perspective, and requires more electronic measurement channels, which can be difficult in terms of power and volume.

\subsubsection{Aspect Ratio}\label{sec:aspectratio}
\begin{figure} [ht]
   \begin{center}
   \begin{tabular}{cc} 
   \includegraphics[height=5cm]{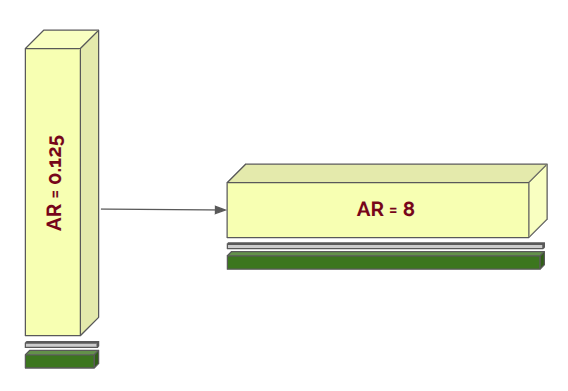}
   \includegraphics[height=5cm]{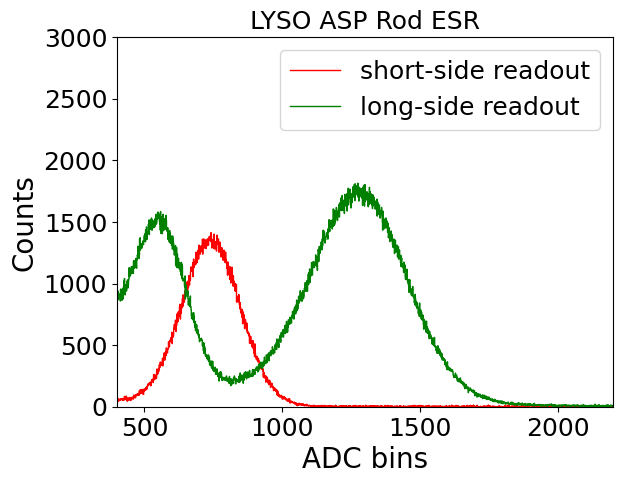}
   \end{tabular}
   \end{center}
   \caption[example]
   { \label{fig:readout}
   Improvement in light collection with an increase in readout face area/aspect ratio (AR). The yellow shapes represent the crystals, the green shapes represent the SiPMs, and the interface represents the optical coupling. The increased light collection is evident in the right plot with the shift in the location of the peaks, with focus on the last peak that corresponds to the 59.5 keV line from $^{241}$Am.}
   \end{figure}

To study the influence of aspect ratio (AR), we tested the LYSO and GAGG rod readout from the short side (5 $\times$ 5 mm) and the long side (5 $\times$ 40 mm). The long-side readout increased the pulse height by up to 72\% for LYSO exposed to the 59.5 keV peak ($^{241}$Am, Figure \ref{fig:readout}), and up to 268\% for GAGG exposed to the same energy. This corresponds to a significant improvement in light collection. However, it also introduced less uniform light collection, which slightly affected the energy resolution. 
This agrees with a prior study\cite{2018PMB....63k5011C} of an LGSO:Ce crystal with dimensions 2.9 $\times$ 2.9 $\times$ 20 mm$^3$ exposed to 511 keV, where $>$90\% of scintillated photons were collected from the long-side readout face compared to the $\sim$40\% collected from the short-side readout face.  

However, the minimal impact of increased light collection on the eventual energy resolution is seen in our results. The long-side readout improved by 5 percentage points for GAGG and 1 percentage point for LYSO, indicating that the increased aspect ratio greatly increases light collection but suffers more from non-uniformity in the light collection.

   \begin{figure} [ht]
   \begin{center}
   \begin{tabular}{c} 
   \includegraphics[width=8cm]{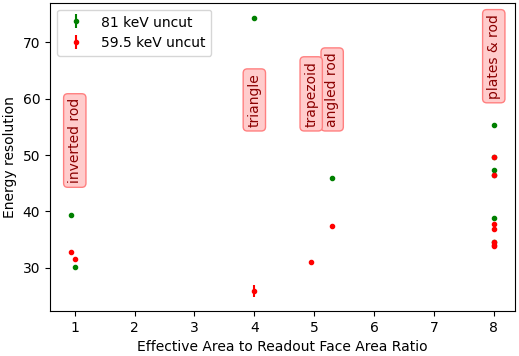}
   \end{tabular}
   \end{center}
   \caption[example]
   { \label{fig:ratio}
   Generally degrading energy resolution with the ratio of the radiation effective area to the readout face area. The inverted rods were read out on the flat long side (5 $\times$ 40 mm face), the angled rod has a 45$^{\circ}$ angle at the short readout face, and "rod" are the normal rods read out on the flat short side (5 $\times$ 5 mm). Ideally, our optimal configuration would be at the lower middle to right corner, which represents a minimized resolution, a large radiation-exposed area, and a small to moderate readout face area.}
   \end{figure}

   \begin{figure} [ht]
   \begin{center}
   \begin{tabular}{cc} 
   \includegraphics[width=5.5cm]{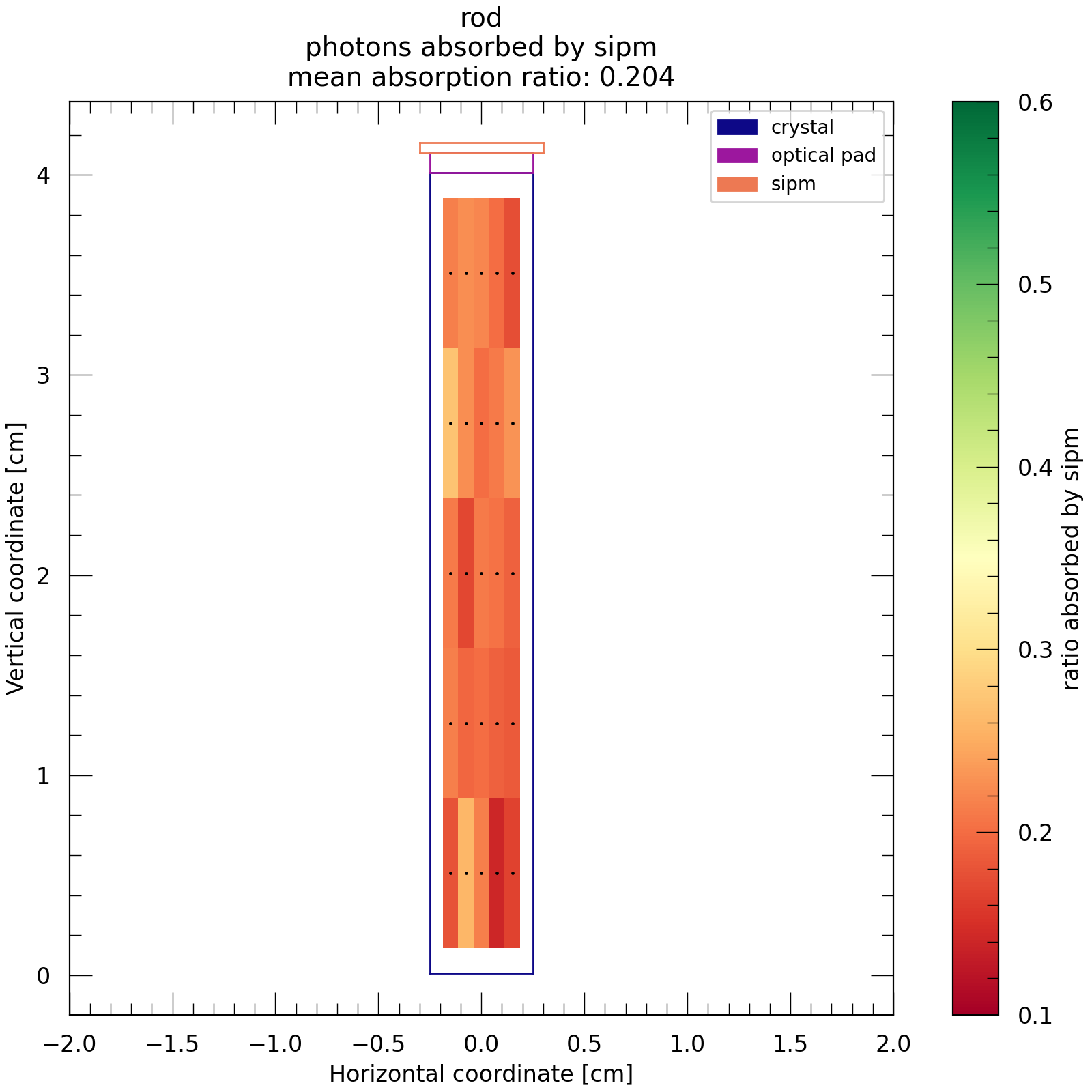}
   \includegraphics[width=5.5cm]{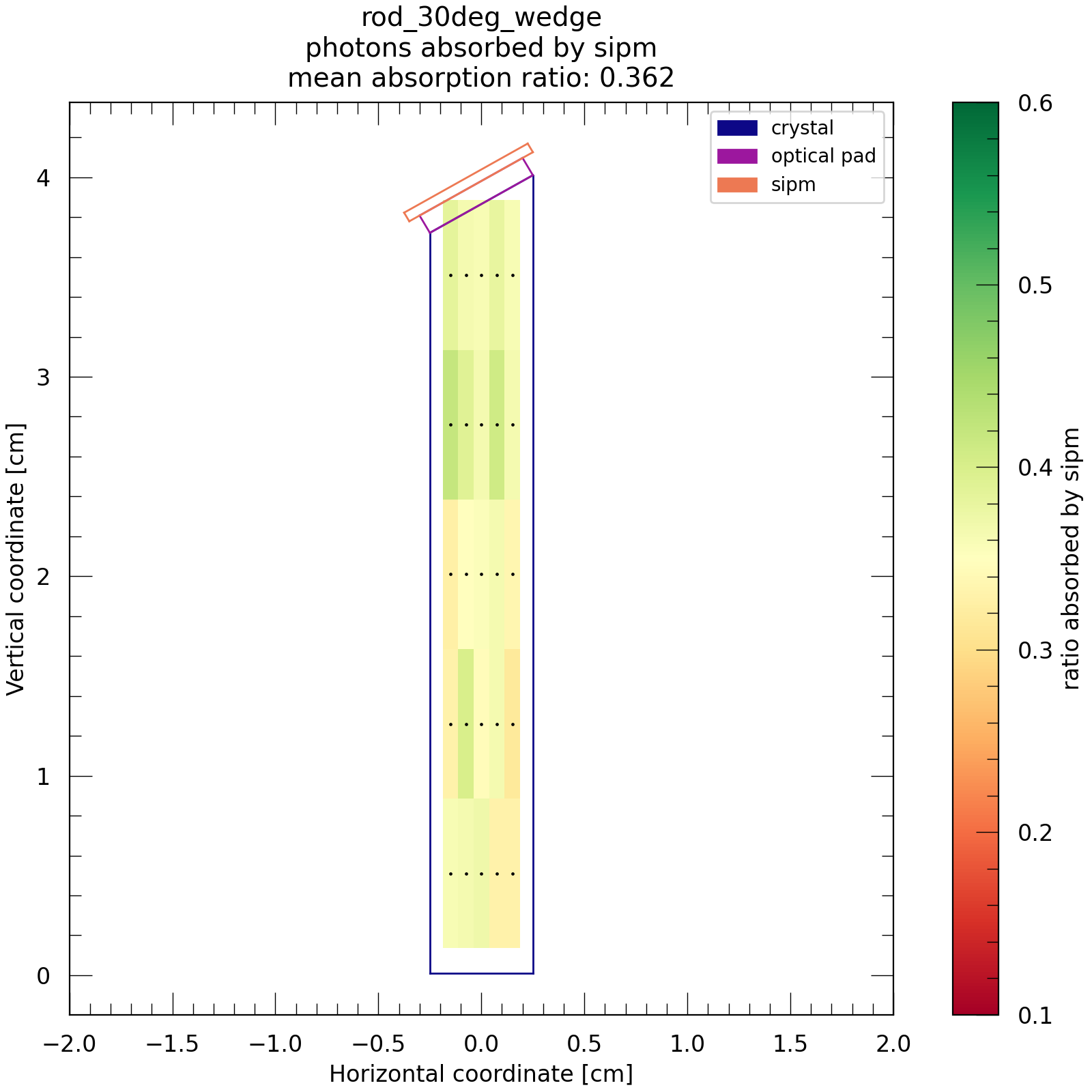}
   \includegraphics[width=5.5cm]{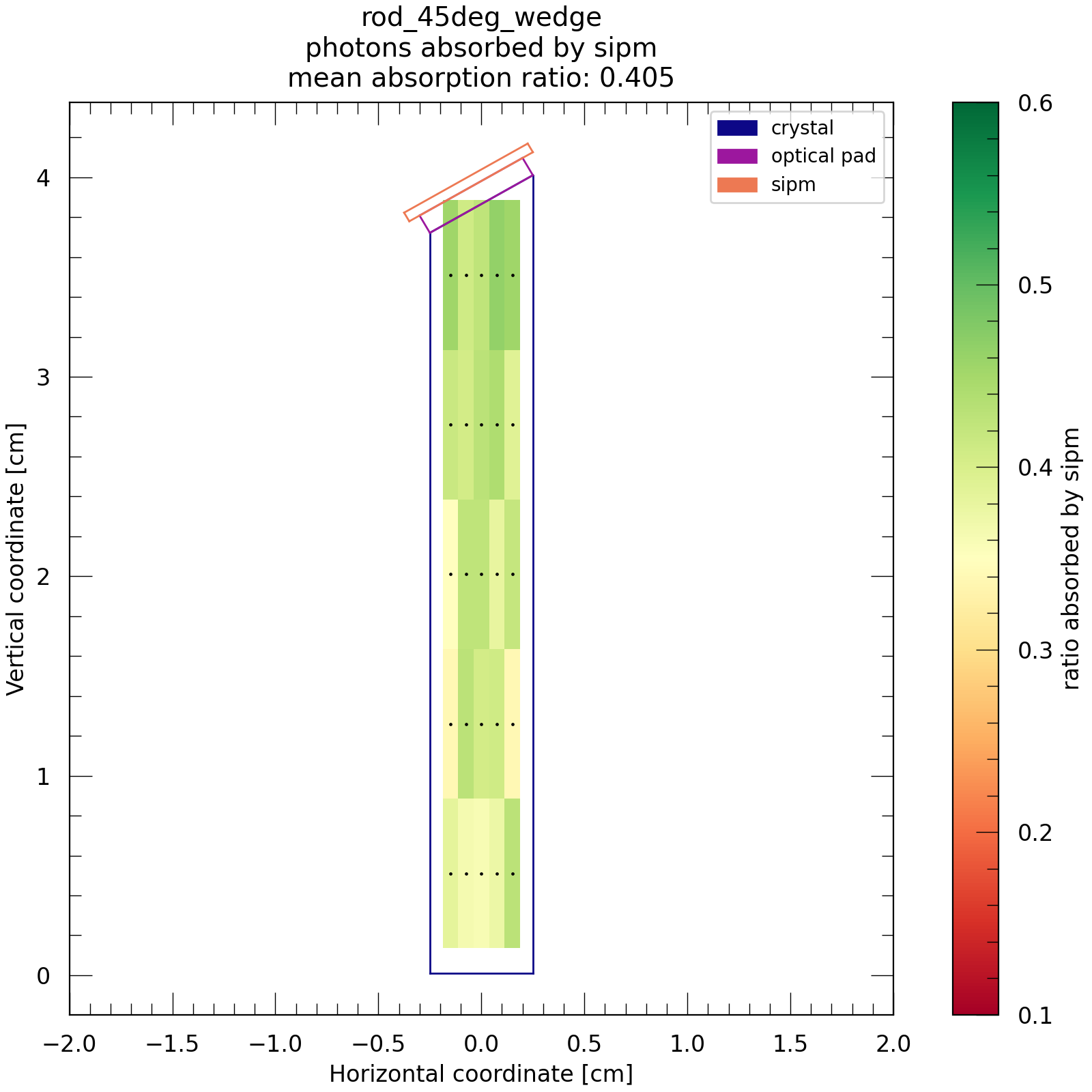}
   \end{tabular}
   \end{center}
   \caption[example]
   { \label{fig:angle}
   \texttt{pvtrace} simulations showing improved light collection at the SiPM by increasing the angle of the readout face (\textit{left to right:} 0$^{\circ}$, 30$^{\circ}$, 45$^{\circ}$) for a rod. The readout face is shown at the top of the rods. The mean ratio of photons absorbed by the SiPM is 0.204 for the flat face, 0.362 for the 30$^{\circ}$ angled face, and 0.405 for the 45$^{\circ}$ angled face.}
   \end{figure}

\subsubsection{Ratio of Radiation-Exposed Area to Readout Face Area}
To evaluate the trade-off between light collection and radiation-exposed area, we computed the ratio of the radiation-exposed area to readout face area. As shown in Figure \ref{fig:ratio}, the energy resolution tends to improve with decreasing ratio, peaking at a value of $\sim$4, which corresponds to the triangle-shaped LYSO crystal. However, the results across energies (59.5 keV and 81 keV) showed significant variability, suggesting that this metric alone is insufficient and needs further investigation. The trapezoid comes close to a good balance at a ratio of 5, exhibiting good resolution, a large effective area, and moderate readout face area.

\subsubsection{Angled Readout Face}
The results show that self-absorption of the scintillating crystals has a substantial impact on the number of photons measured by the SiPMs.

Simulations with \verb|pvtrace| (Section \ref{sec:simulation}) were used to test rods with readout face angles of 0$^\circ$, 30$^\circ$, and 45$^\circ$. The mean photon collection efficiencies at the SiPM were 0.204, 0.362, and 0.405, respectively. The 45$^\circ$-angled face yields the best light collection rate, which is nearly double the flat face value (Figure \ref{fig:angle}). This improvement is likely due to reduced internal reflection, as the photons impinge on the readout face at an angle less than the critical angle, as well as shorter average photon path lengths within the crystal.

\begin{figure} [t]
   \begin{center}
   \begin{tabular}{cc} 
   \includegraphics[height=5cm]{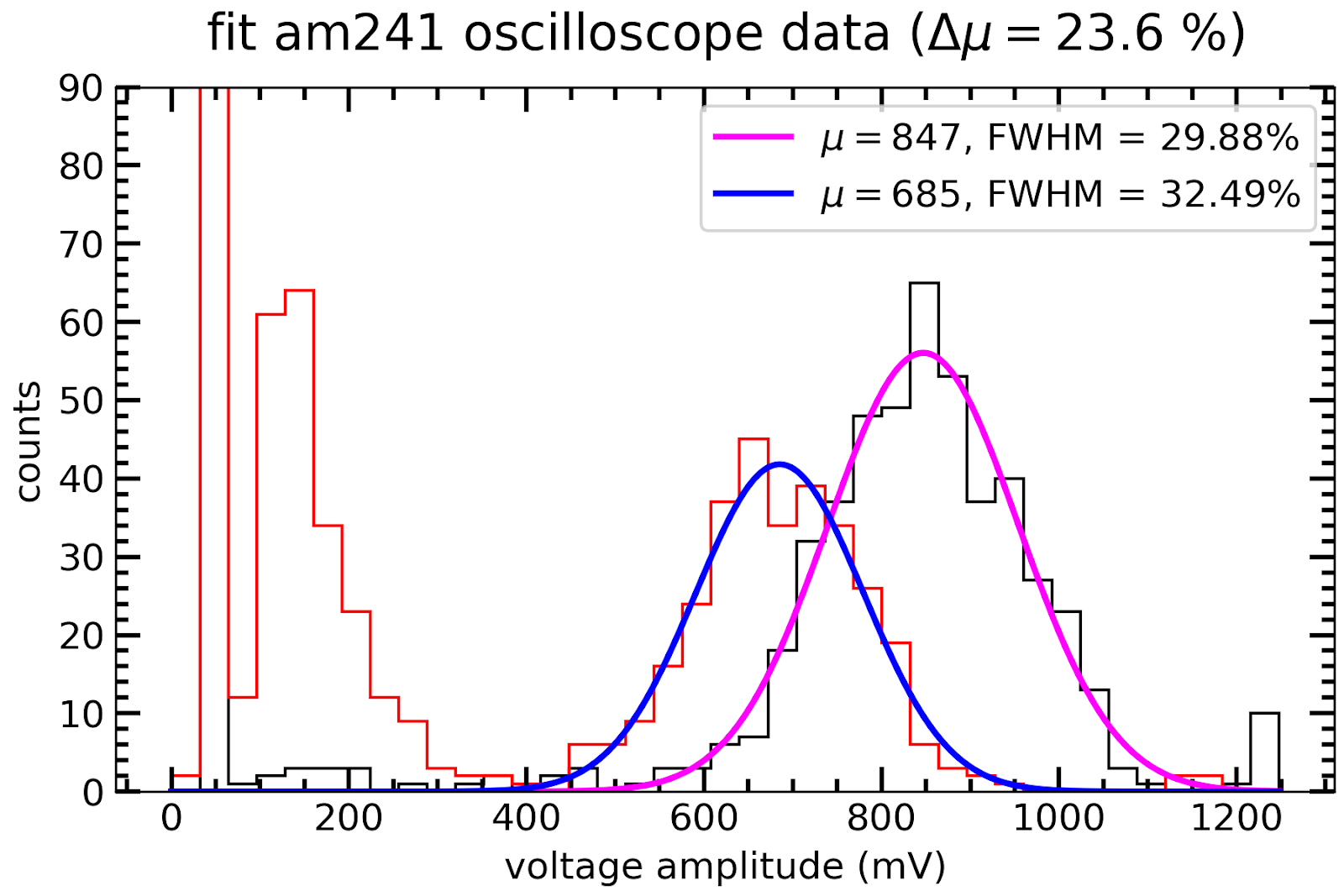}
   \end{tabular}
   \end{center}
   \caption[example]
   { \label{fig:cut}
   The $^{241}$Am 59.5 keV energy peak for LYSO rod with a flat and angled readout face. The black spectrum corresponds to the angled readout, and the red represents the flat readout face. The plot demonstrates an improvement in pulse amplitude and energy resolution with a flat readout face having a FWHM of 33\% (blue fit) and the 45$^\circ$ angled face with an FWHM of 30\% (magenta fit).}
   \end{figure}

Lab results are consistent with the simulations: the LYSO rod with a 45$^\circ$-cut showed a 20\% increase in pulse height and a 3 percentage point improvement in energy resolution at the 59.5 keV peak (Figure \ref{fig:cut}). 

   \begin{figure} [t]
   \begin{center}
   \begin{tabular}{cc} 
   \includegraphics[width=5.5cm]{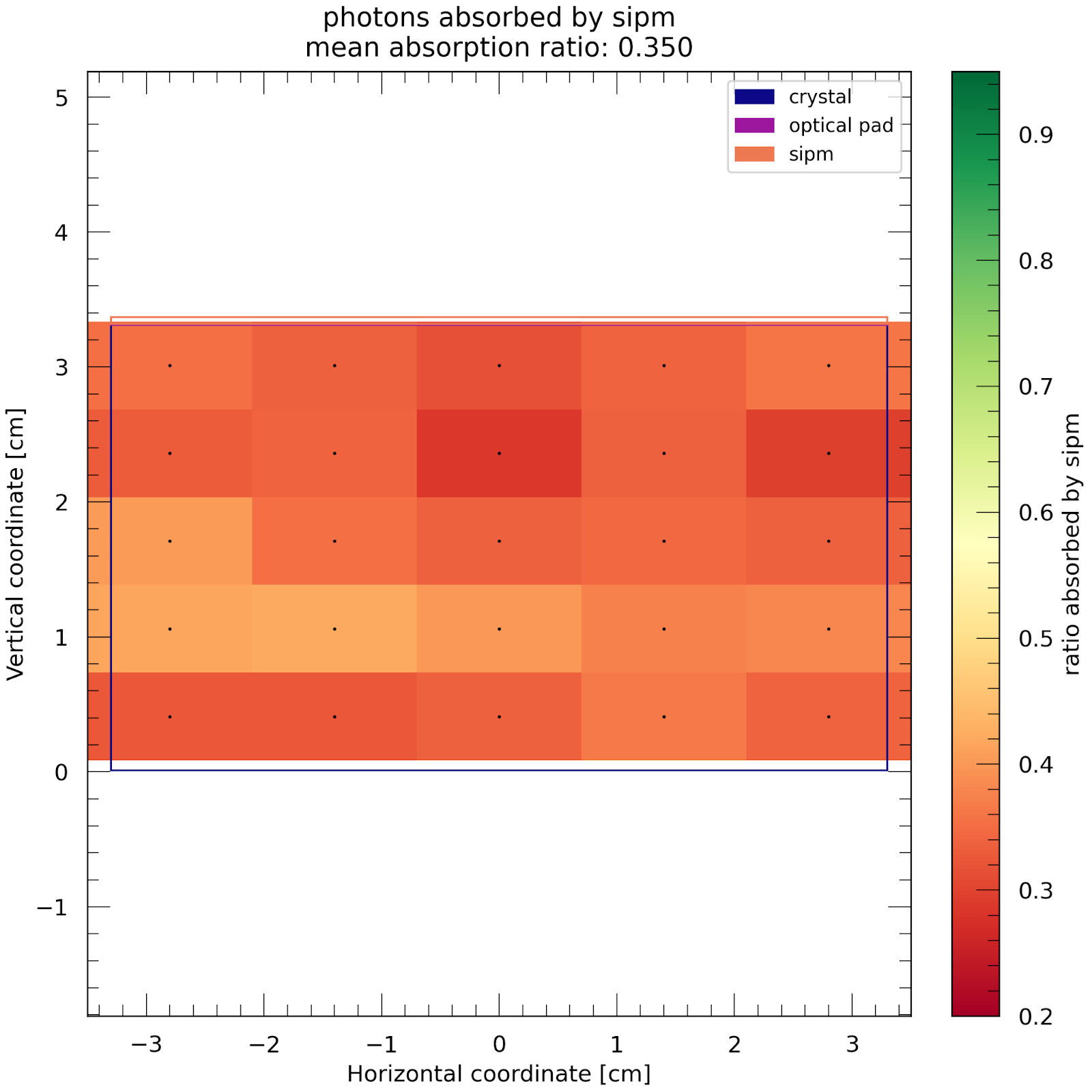}
   \includegraphics[width=5.5cm]{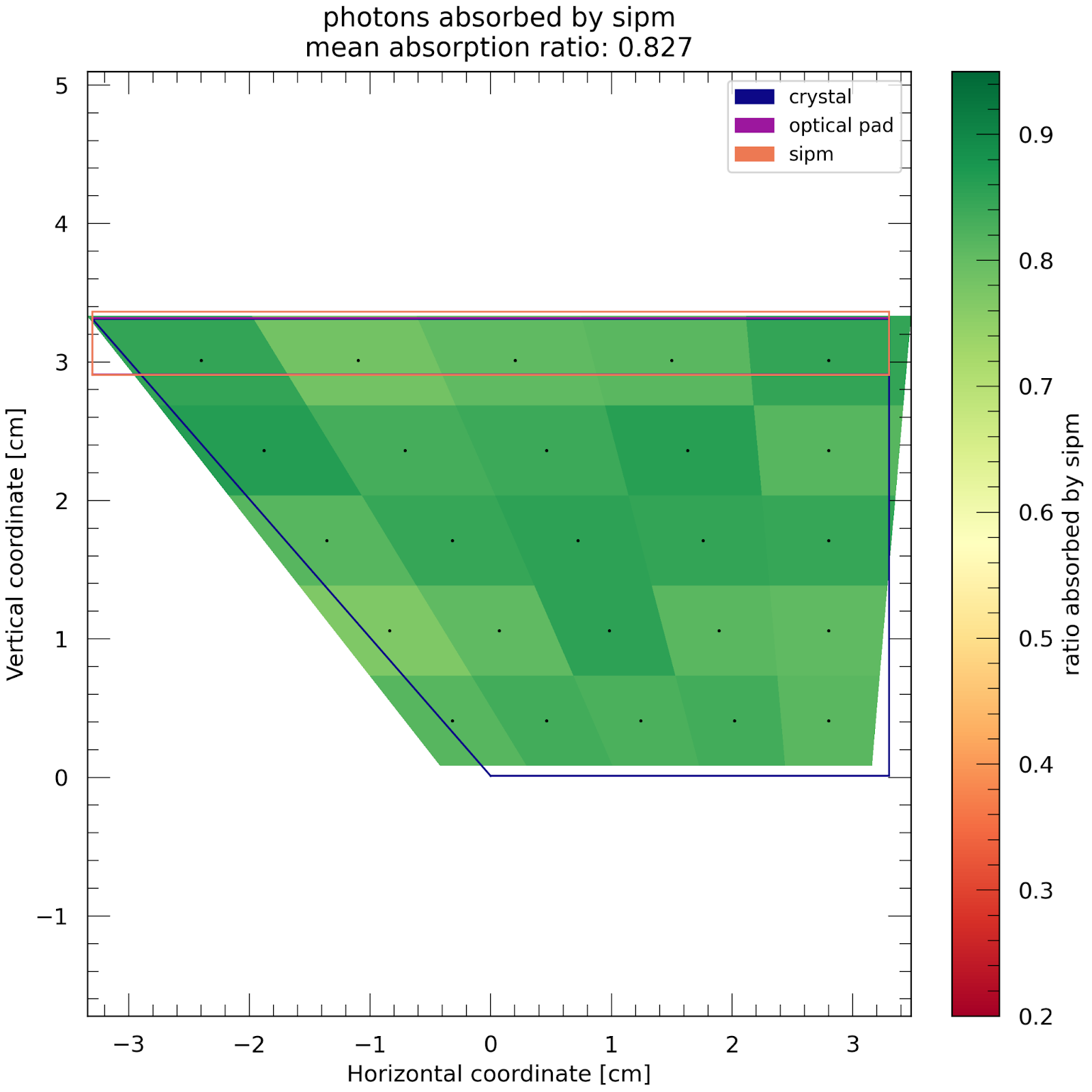}
   \includegraphics[width=5.5cm]{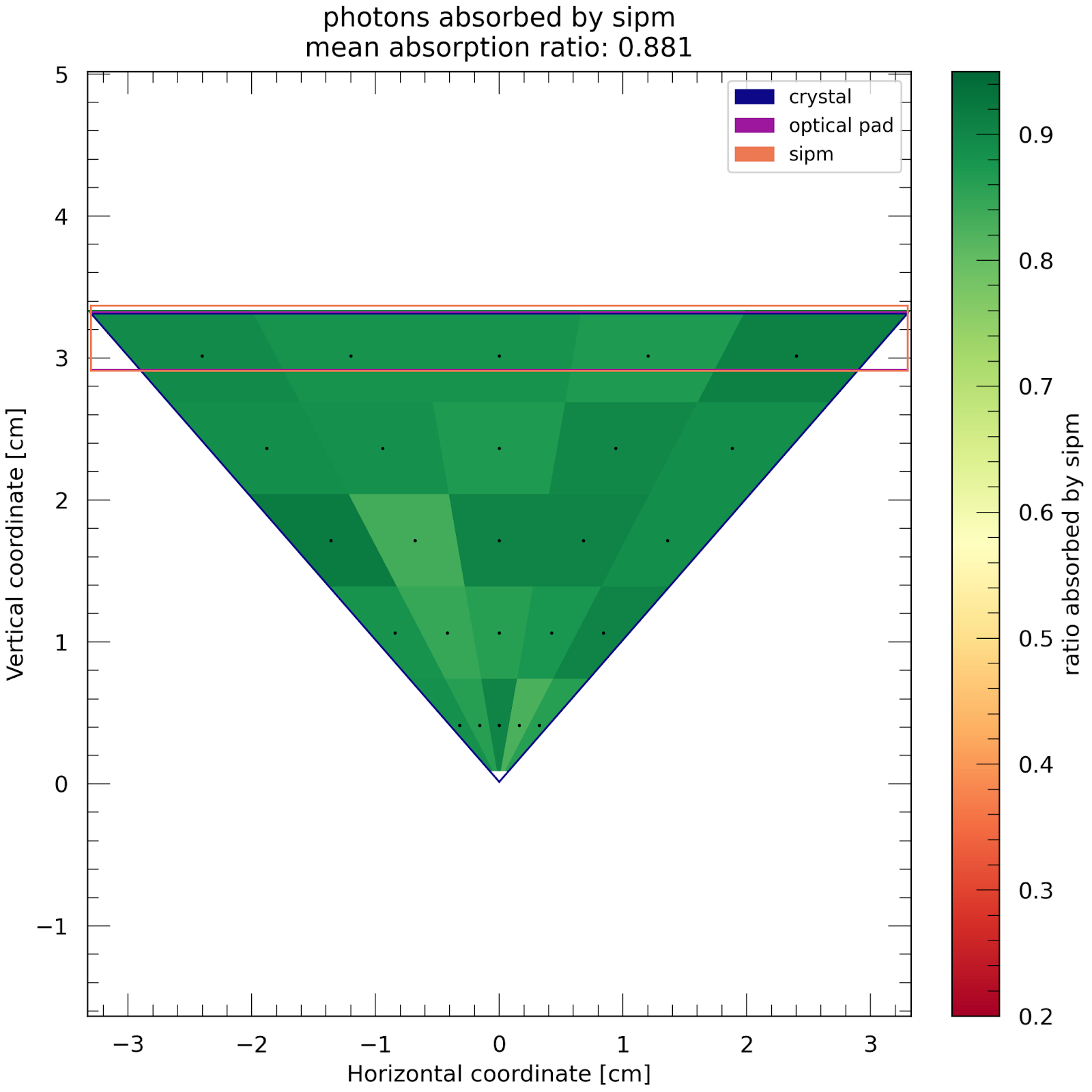}
   \end{tabular}
   \end{center}
   \caption[example]
   { \label{fig:platecut}
   \texttt{pvtrace} simulations of LYSO plates showing an improvement in light collection at the SiPMs with an increase in cuts at the plate edges and the readout face. The readout face is at the top, and the angled readout face is represented by the red rectangle drawn at the top. The mean absorption ratio at the SiPM is 0.35 for the uncut plate, 0.83 for the trapezoid with angled readout face, and 0.88 for the triangle with angled readout face.}
   \end{figure}

\begin{figure} [ht]
   \begin{center}
   \begin{tabular}{cc} 
   \includegraphics[width=13cm]{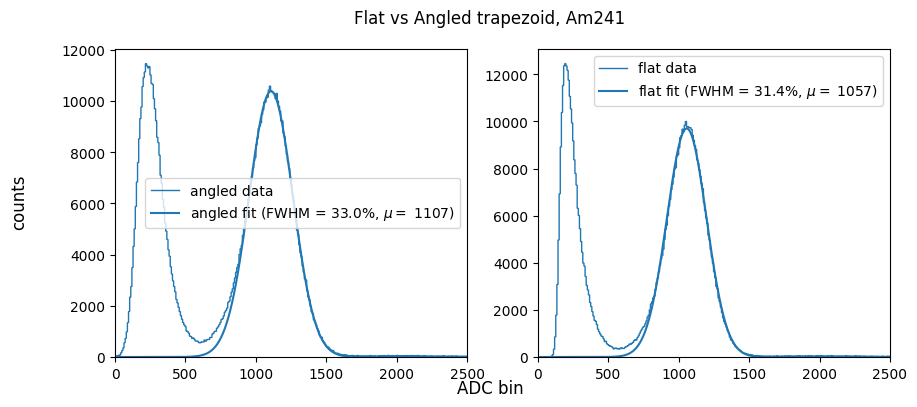}
   \end{tabular}
   \end{center}
   \caption[example]
   { \label{fig:flatangletrap}
   The $^{241}$Am 59.5 keV energy peak for LYSO trapezoid with angled (\textit{left}) and flat (\textit{right}) readout faces. The angled readout has an improved pulse amplitude and energy resolution.}
   \end{figure}

To extend this to larger-area detectors, we extended the same experiment for the plate geometries. The progression of simulations was from the flat readout face plate configuration (20 mm $\times$ 5 mm) to a plate with a 45$^\circ$ angled readout face, to a trapezoid with an angled readout face, to a triangle with an angled readout face. The simulations show a rise in light collection from 0.35 (flat plate) to 0.881 (triangular plate with angled readout), as shown in Figure \ref{fig:platecut}. Although the triangular plate had the highest light collection in the simulations, lab tests yielded mixed results: a best-case resolution of 25.86\% (59.5 keV), but also a worst-case resolution of 74.32\% (81 keV), indicating inconsistent performance. This is in part due to the difficulty in wrapping ESR tightly around the triangle edges, without it slipping off, and effectively covering the angled readout face region. Improved light collection may introduce non-uniform photon arrival, which can degrade resolution, similar to long-side vs. short-side readout trends discussed in Sec. \ref{sec:aspectratio}. This is further evidenced in our comparison between lab tests of flat and angle trapezoid (Figure \ref{fig:flatangletrap}). The flat readout has a lower light collection with peak mean, $\mu$=1057, but better energy resolution due to the uniformity in the light collected.

   \begin{figure} [H]
   \begin{center}
   \begin{tabular}{c} 
   \includegraphics[height=2.5cm]{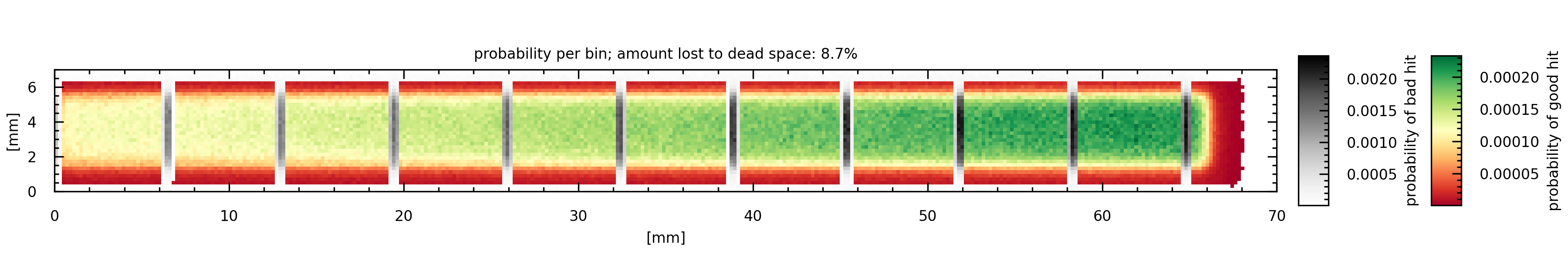}\\
   \includegraphics[height=3.4cm]{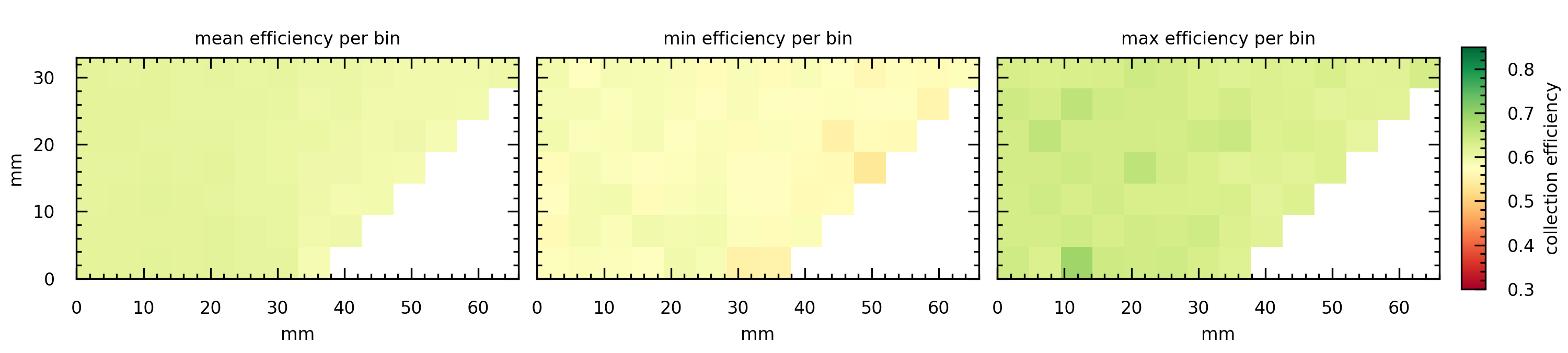}\\
   \includegraphics[height=2.5cm]{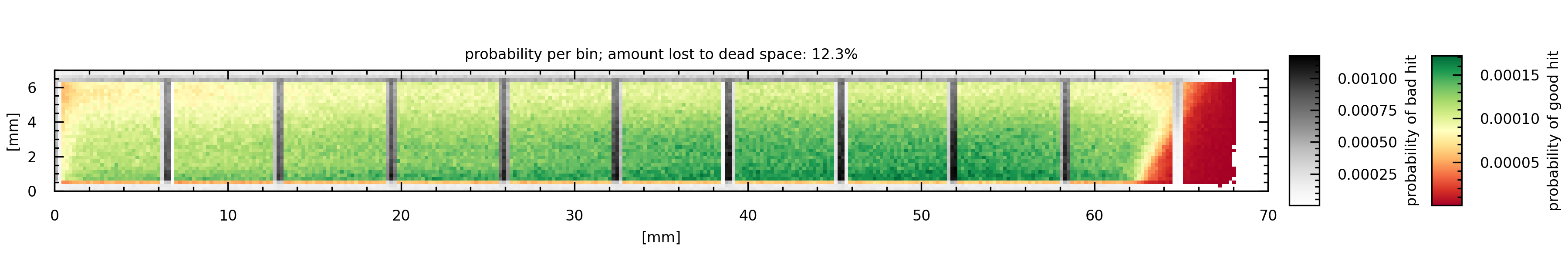}\\
   \includegraphics[height=3.4cm]{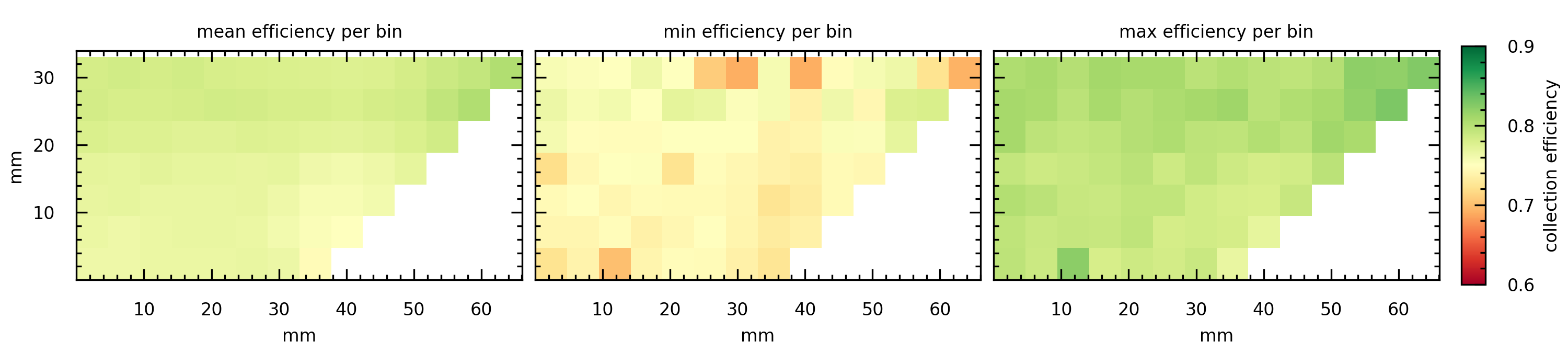}
   \end{tabular}
   \end{center}
   \caption[example]
   { \label{fig:trapflatangle}
   From top to bottom: Geant4 simulations showing the spatial distribution of photons incident on the ten SiPMs coupled to a flat trapezoid, the spatial distribution of the light collection efficiency (LCE) for the flat trapezoid, the 10 SiPMs coupled to an angled-readout trapezoid, and the spatial distribution of the LCE for the angled-readout trapezoid. The readout face is at the top (the longest face of the trapezoid). The plot shows a higher probability of photons hitting the SiPMs with the angled readout, especially for photons scintillated close to the readout face.}
   \end{figure}

\begin{figure} [t]
   \begin{center}
   \begin{tabular}{cc} 
   \includegraphics[height=5.5cm]{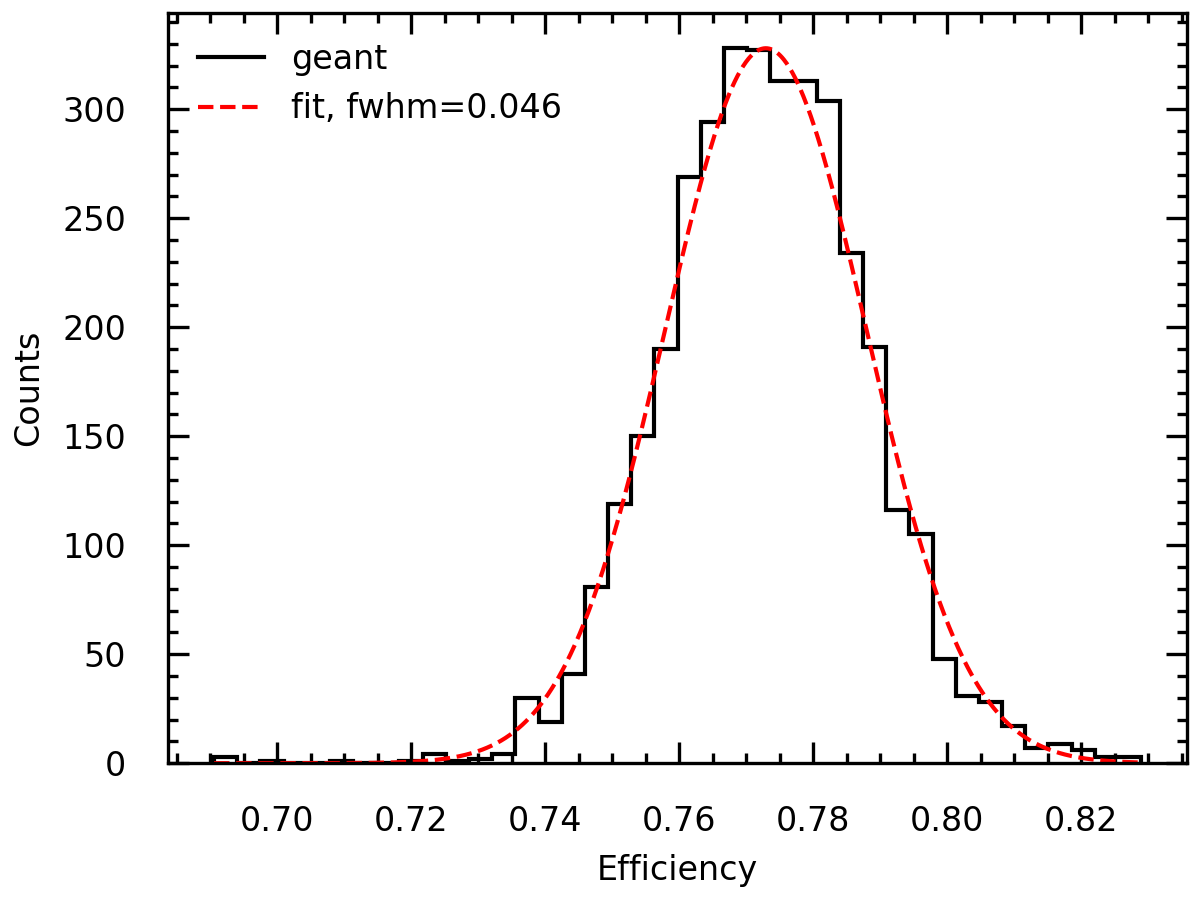}
   \includegraphics[height=5.5cm]{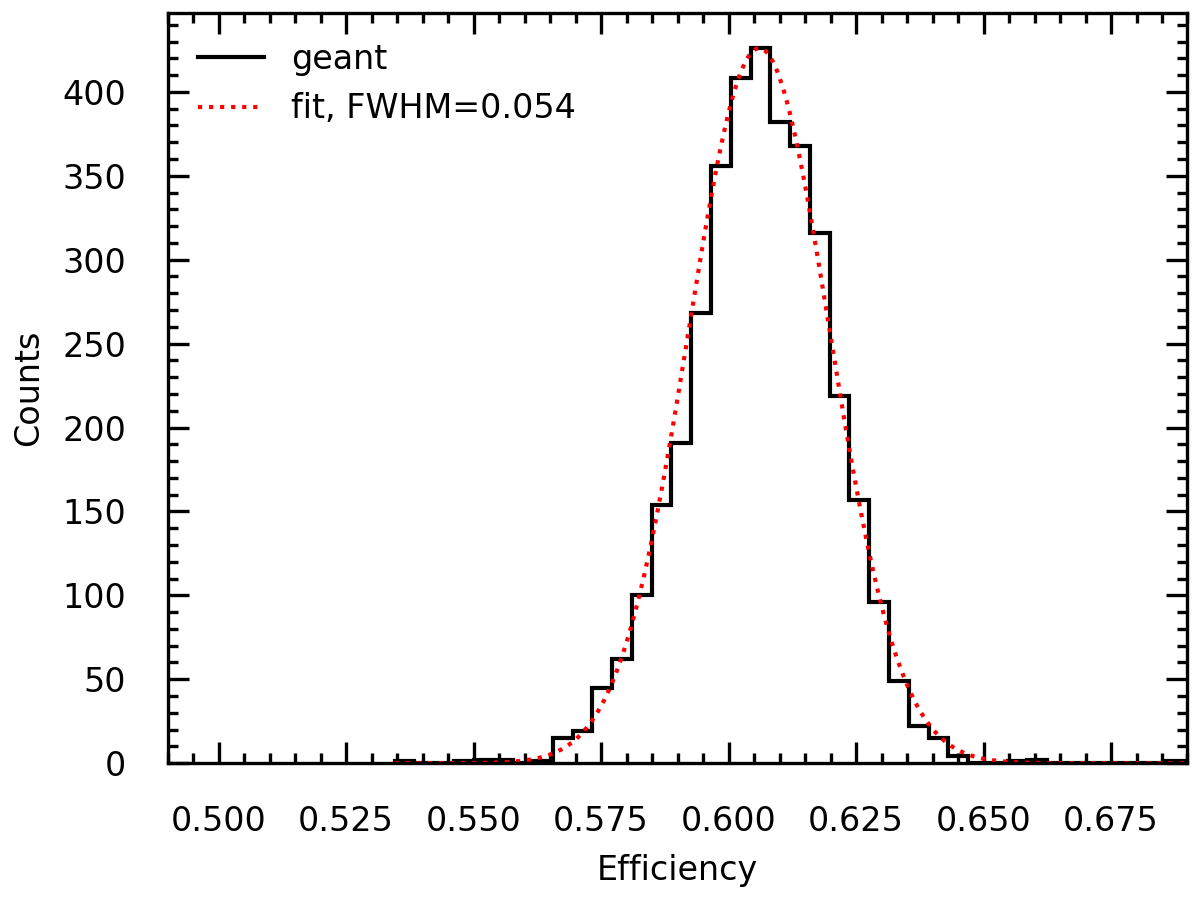}
   \end{tabular}
   \end{center}
   \caption[example]
   { \label{fig:trapcut}
   Geant4 simulation of the distribution of the photon collection efficiency of 10,000 X-rays fired at the angled-readout (\textit{left}) and flat (\textit{right}) trapezoids. These plots show the spatially integrated LCEs of the results shown in Figure~\ref{fig:trapflatangle}. The flat and angled trapezoid has a Gaussian distribution. }
   \end{figure}

We also studied some of these geometries with Geant4 simulation by firing 10,000 X-rays randomly in the trapezoid plane, assuming a \SI{50}{\per\centi\meter} absorption coefficient for LYSO\cite{2006NIMPA.564..506V}. The results confirm that an angled readout face improves light collection efficiency, as shown by the  higher maximum and mean efficiency per bin in the last two panels of Figure \ref{fig:trapflatangle}, compared to the first two panels. The improvement is further supported by the shift of the photon distribution peak to a higher light collection efficiency value in the left panel of Figure \ref{fig:trapcut}.

Although the angled readout improved light collection, the flat readout trapezoidal design offered a balance between mechanical stability and performance, achieving a collection ratio of 0.827 in \verb|pvtrace| simulation and an energy resolution of 31.4\% in lab measurements. This geometry was determined to be sufficient for our work.

The improvements across different geometries tested suggest that angled cuts reduce photons lost due to self-absorption by reducing total internal reflection at the readout interface. The simulations confirmed that the angled readout improves total light collection, but the uniformity of photon arrival is a critical factor that can counteract those gains, emphasizing the trade-off between light quantity and quality.

\subsection{Large Area vs Dark Count Tradeoff}
Trapezoids have provided us with an excellent balance of good energy resolution and increased effective area, although they require ten SiPMs rather than the six needed for our plates and the one needed for our rods. 

First, we calculated the dark count amplitude at specific temperatures, and they were all consistent with the measured values at these temperatures. We also measured the dark counts variation with temperature from -20$^{\circ}$C to 23$^{\circ}$C, and the mean pulse amplitude (mV) increases with temperature, as expected. 

Then, we compared the energy spectrum of $^{241}$Am and the energy resolution at 59.5 keV acquired with the LYSO ASP rod and trapezoid. Although the trapezoid arrangement with 10 SiPMs had ten times more dark counts than the rod with a single SiPM, the trapezoid's resolution improved by 4 percentage points (36\% to 32\%, Figure \ref{fig:darkcount} in Appendix). The improvement in the photon collection by the trapezoid outweighed the influence of dark counts on the energy resolution.

\subsection{Optical Coupling}\label{padresults}
Since the optical interface between the crystal and the SiPM greatly influences the photons reaching the sensor, we investigated the impact of optical pad thickness on light collection.

\begin{figure} [t]
   \begin{center}
   \begin{tabular}{c} 
   \includegraphics[height=5.5cm]{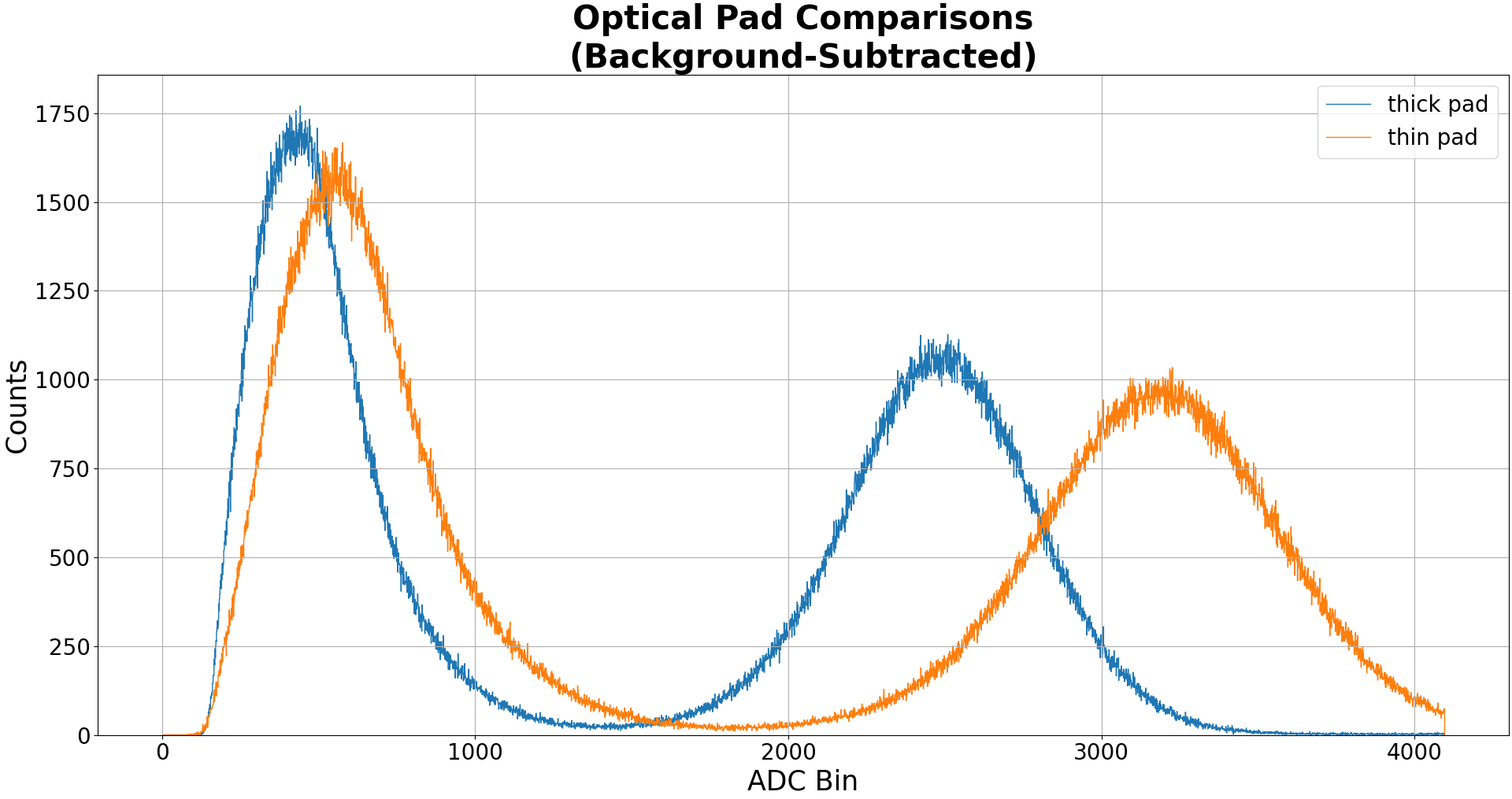}
   \end{tabular}
   \end{center}
   \caption[example]
   { \label{fig:padsplot}
   $^{241}$Am data collected using the single-SiPM setup. The increase in total counts from the thin optical pad to the thick optical pad is directly correlated to an increase in light collection.}
   \end{figure}

The claim that a thinner optical pad leads to greater light collection was tested using a single-SiPM setup, where a rectangular LYSO rod (5 mm $\times$ 5 mm $\times$ 40 mm) was placed with its small end on a singular SiPM. This SiPM was coupled to the rod using both the 1 mm thick EJ-560 optical pad as well as the Sylgard 184 $\sim$0.25 mm thin optical pad. Figure \ref{fig:padsplot} shows the collected $^{241}$Am spectra for both of these setups. There was a 16\% increase in light collection for the thin pad as compared to the thick pad.

\section{CONCLUSION}

A series of experiments and simulations were conducted to optimize the performance of IMPISH, a hard X-ray spectrometer. By comparing the crystal types, reflectors, geometries, surface properties, and readout configurations, some key findings emerged. LYSO was chosen over GAGG, due to its superior spectral match with the SiPM sensitivity and faster decay time, which outweighs its lower and non-linearity in the intrinsic light yield.

Crystals with all sides polished yielded slightly better energy resolution than partially polished ones, and Teflon (Lambertian) reflectors performed better than ESR (specular). Geometric tests show that rod-shaped crystals provided better resolution, but plate and trapezoidal geometries were preferred because of their larger effective area and reduced readout channels. Increasing the readout face area and introducing angled cuts significantly improved light collection efficiency.

Additional improvements were achieved through improved optical coupling, with a custom 0.25 mm Sylgard 184 pad that increased light collection by 16\% over a standard 1 mm Eljen pad.  Although the trapezoid geometry required ten SiPMs (higher dark count rates), it still yielded a better energy resolution than the single-SiPM rod, showing that the enhanced light collection outweighed the increased dark current. Simulations with Geant4 and \verb|pvtrace| supported our experimental findings, especially with respect to internal absorption, total internal reflection, and uniformity in light collection. Overall, the LYSO trapezoid crystal provides better performance for the IMPISH detector design by balancing light yield, energy resolution, and mechanical constraints.

\appendix    

\section{EXTRA PLOTS}

\subsection{Comparison of Energy Spectra}
   
   \begin{figure} [H]
   \begin{center}
   \begin{tabular}{cc} 
   \includegraphics[height=5cm]{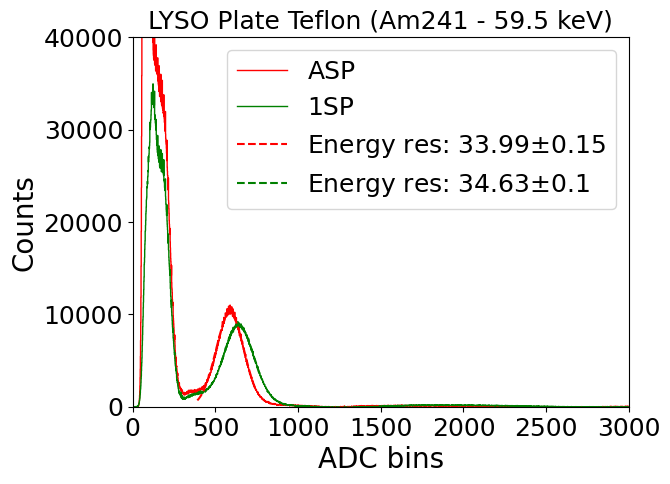}
   \includegraphics[height=5cm]{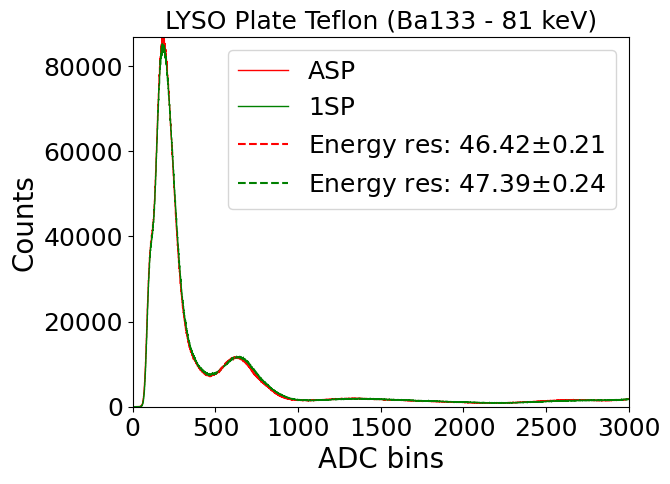}
   \end{tabular}
   \end{center}
   \caption[example]
   { \label{fig:rough} Energy spectra for $^{241}$Am and $^{133}$Ba showing all sides polished (ASP) vs. one side polished (1SP) LYSO plate wrapped with Teflon reflector. ASP energy resolution in \% is similar (\textit{left}) or slightly better than 1SP crystals (\textit{right}).}
   \end{figure}

   \begin{figure} [H]
   \begin{center}
   \begin{tabular}{cc} 
   \includegraphics[height=5cm]{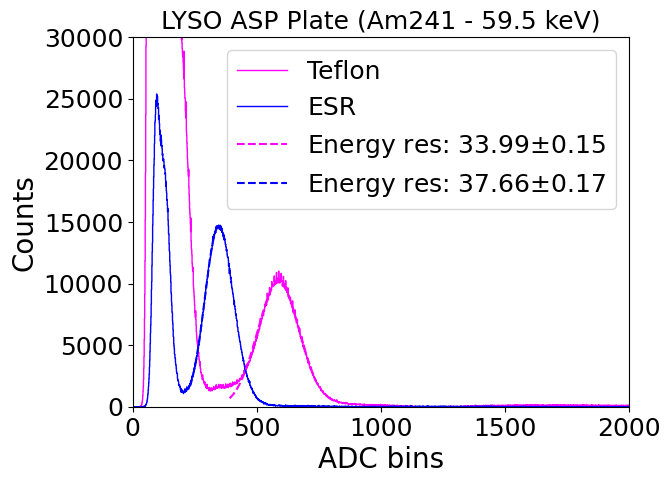}
   \includegraphics[height=5cm]{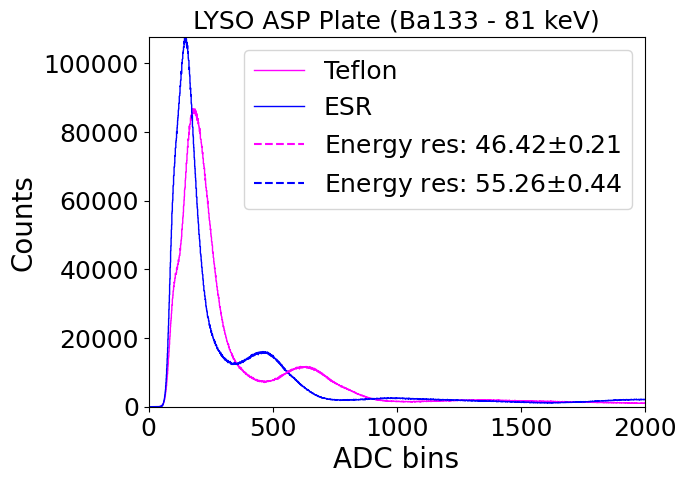}
   \end{tabular}
   \end{center}
   \caption[example]
   { \label{fig:reflect}
   Energy spectra for $^{241}$Am and $^{133}$Ba showing Teflon (magenta) diffuse reflection yielding better energy resolution (in \%) and light collection compared to ESR (blue) specular reflection.}
   \end{figure}

   \begin{figure} [H]
   \begin{center}
   \begin{tabular}{c} 
   \includegraphics[height=5cm]{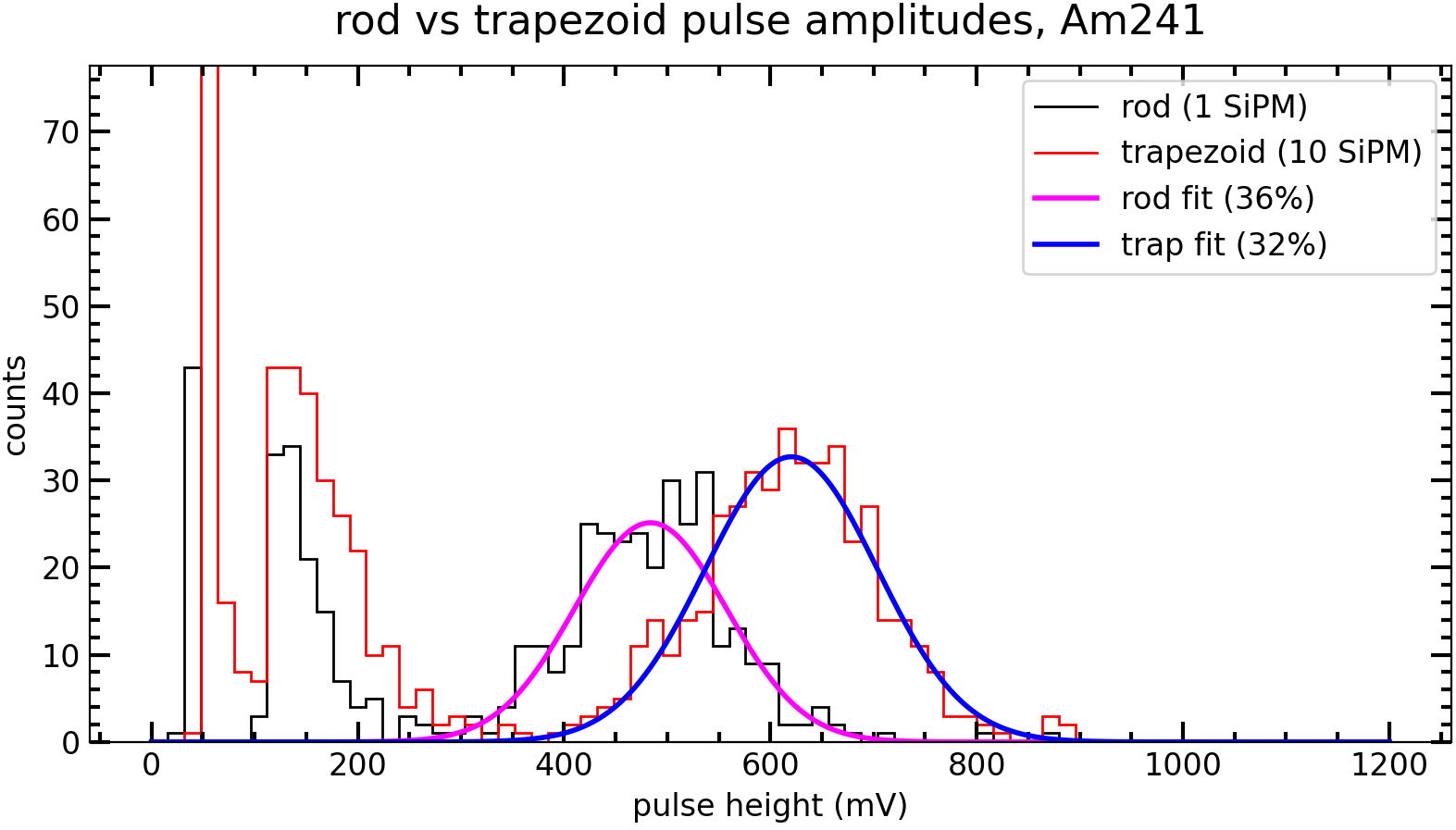}
   \end{tabular}
   \end{center}
   \caption[example]
   { \label{fig:darkcount} 
Improved energy resolution of the $^{241}$Am 59.5 keV peak from rod (36\%), coupled to 1 SiPM, to trapezoid (32\%), coupled to 10 SiPMs, despite the 10-times increase in the SiPM dark count.}
   \end{figure}

\acknowledgments 

This research is funded by the NASA Low-Cost Access to Space grant 80NSSC24M0030. We also thank the staff of the Earth Science Lab at the University of Minnesota, Twin Cities, for their assistance.

\bibliography{report} 

\begin{thebibliography}{10}

\bibitem{2022SPIE12181E..3MS}
{Setterberg}, W., {Glesener}, L., {Gebre-Egziabher}, D., {Sample}, J.~G., {Smith}, D.~M., {Caspi}, A., {Faulkner}, A., {Clemmer}, L., {Hildebrandt}, K., {Skinner}, E., {Greathouse}, A., {Kozic}, T., {Wieber}, M., {Savadogo}, M., {Nightingale}, M., and {Knuth}, T., ``{Geant4 modeling of a cerium bromide scintillator detector for the IMPRESS CubeSat mission},'' in [{\em Space Telescopes and Instrumentation 2022: Ultraviolet to Gamma Ray}{\nolinebreak\hspace{0.1em}]},  {den Herder}, J.-W.~A., {Nikzad}, S., and {Nakazawa}, K., eds., {\em Society of Photo-Optical Instrumentation Engineers (SPIE) Conference Series} {\bf 12181},  121813M (Aug. 2022).

\bibitem{2021PhDT.........2K}
{Knuth}, T.~J., {\em {Studying Particle Acceleration in Solar Flares via Subsecond X-Ray Spikes: Analysis and Instrumentation}}, PhD thesis, University of Minnesota, Twin Cities (Jan. 2021).

\bibitem{2020ApJ...903...63K}
{Knuth}, T. and {Glesener}, L., ``{Subsecond Spikes in Fermi GBM X-Ray Flux as a Probe for Solar Flare Particle Acceleration},'' {\em The Astrophysical Journal}~{\bf 903},  63 (Nov. 2020).

\bibitem{reed2025the}
Masek, R.~B., Setterberg, W., Williams, P., Clemmer, L., Oseni, D., Oliphant, L., Glesener, L., Gebre-Egziabher, D., Sample, J.~G., Caspi, A., Saint-Hilaire, P., Shih, A., and Smith, D.~M., ``The integrating miniature piggyback for impulsive solar hard x-rays (impish): a low-cost spectrometer for the grips-2 balloon campaign,'' in [{\em Astronomical Applications 2025: UV, X-Ray, and Gamma-Ray Space Instrumentation for Astronomy}{\nolinebreak\hspace{0.1em}]},  SPIE (2025).

\bibitem{smith2002rhessi}
Smith, D.~M., Lin, R., Turin, P., Curtis, D., Primbsch, J., Campbell, R., Abiad, R., Schroeder, P., Cork, C., Hull, E., et~al., ``The rhessi spectrometer,'' {\em Solar Physics}~{\bf 210}(1),  33--60 (2002).

\bibitem{shih2012gamma}
Shih, A.~Y., Lin, R.~P., Hurford, G.~J., Duncan, N.~A., Saint-Hilaire, P., Bain, H.~M., Boggs, S.~E., Zoglauer, A.~C., Smith, D.~M., Tajima, H., et~al., ``The gamma-ray imager/polarimeter for solar flares (grips),'' in [{\em Space Telescopes and Instrumentation 2012: Ultraviolet to Gamma Ray}{\nolinebreak\hspace{0.1em}]},   {\bf 8443},  1297--1311, SPIE (2012).

\bibitem{krucker2020spectrometer}
Krucker, S., Hurford, G.~J., Grimm, O., K{\"o}gl, S., Gr{\"o}belbauer, H.-P., Etesi, L., Casadei, D., Csillaghy, A., Benz, A.~O., Arnold, N.~G., et~al., ``The spectrometer/telescope for imaging x-rays (stix),'' {\em Astronomy \& Astrophysics}~{\bf 642},  A15 (2020).

\bibitem{meegan2009fermi}
Meegan, C., Lichti, G., Bhat, P., Bissaldi, E., Briggs, M.~S., Connaughton, V., Diehl, R., Fishman, G., Greiner, J., Hoover, A.~S., et~al., ``The fermi gamma-ray burst monitor,'' {\em The Astrophysical Journal}~{\bf 702}(1),  791 (2009).

\bibitem{2000rdm..book.....K}
{Knoll}, G.~F.,  [{\em {Radiation detection and measurement}}{\nolinebreak\hspace{0.1em}]}, John \& Wiley Sons Inc, 3~ed. (2000).

\bibitem{2011LanB.21B1...45L}
{Lecoq}, P., ``{Scintillation Detectors for Charged Particles and Photons},'' {\em Landolt B{\"o}rnstein}~{\bf 21B1},  45 (Jan. 2011).

\bibitem{2004ITNS...51.1084P}
{Pidol}, L., {Kahn-Harari}, A., {Viana}, B., {Virey}, E., {Ferrand}, B., {Dorenbos}, P., {Dehaas}, J.~T.~M., and {Vaneijk}, C.~W.~E., ``{High Efficiency of Lutetium Silicate Scintillators, Ce-Doped LPS, and LYSO Crystals},'' {\em IEEE Transactions on Nuclear Science}~{\bf 51},  1084--1087 (June 2004).

\bibitem{2013ITNS...60..988Y}
{Yeom}, J.~Y., {Yamamoto}, S., {Derenzo}, S.~E., {Spanoudaki}, V.~C., {Kamada}, K., {Endo}, T., and {Levin}, C.~S., ``{First Performance Results of Ce:GAGG Scintillation Crystals With Silicon Photomultipliers},'' {\em IEEE Transactions on Nuclear Science}~{\bf 60},  988--992 (Apr. 2013).

\bibitem{2013PMB....58.2185R}
{Roncali}, E. and {Cherry}, S.~R., ``{Simulation of light transport in scintillators based on 3D characterization of crystal surfaces},'' {\em Physics in Medicine and Biology}~{\bf 58},  2185 (Apr. 2013).

\bibitem{1999NIMPA.437..374H}
{Huber}, J.~S., {Moses}, W.~W., {Andreaco}, M.~S., {Loope}, M., {Melcher}, C.~L., and {Nutt}, R., ``{Geometry and surface treatment dependence of the light collection from LSO crystals},'' {\em Nuclear Instruments and Methods in Physics Research A}~{\bf 437},  374--380 (Nov. 1999).

\bibitem{2012ITNS...59.2340P}
{Pauwels}, K., {Auffray}, E., {Gundacker}, S., {Knapitsch}, A., and {Lecoq}, P., ``{Effect of Aspect Ratio on the Light Output of Scintillators},'' {\em IEEE Transactions on Nuclear Science}~{\bf 59},  2340--2345 (Oct. 2012).

\bibitem{2009ITNS...56.3800C}
{Chewpraditkul}, W., {Swiderski}, L., {Moszynski}, M., {Szczesniak}, T., {Syntfeld-Kazuch}, A., {Wanarak}, C., and {Limsuwan}, P., ``{Scintillation Properties of LuAG:Ce, YAG:Ce and LYSO:Ce Crystals for Gamma-Ray Detection},'' {\em IEEE Transactions on Nuclear Science}~{\bf 56},  3800--3805 (Dec. 2009).

\bibitem{gektin2017inorganic}
Gektin, A. and Korzhik, M.,  [{\em Inorganic scintillators for detector systems}{\nolinebreak\hspace{0.1em}]}, Springer (2017).

\bibitem{2017isds.book.....L}
{Lecoq}, P., {Gektin}, A., and {Korzhik}, M.,  [{\em {Inorganic Scintillators for Detector Systems: Physical Principles and Crystal Engineering}}{\nolinebreak\hspace{0.1em}]}, Springer (2017).

\bibitem{chewpraditkul2009scintillation}
Chewpraditkul, W., Swiderski, L., Moszynski, M., Szczesniak, T., Syntfeld-Kazuch, A., Wanarak, C., and Limsuwan, P., ``Scintillation properties of luag: Ce, yag: Ce and lyso: Ce crystals for gamma-ray detection,'' {\em IEEE Transactions on Nuclear Science}~{\bf 56}(6),  3800--3805 (2009).

\bibitem{2009ITNS...56.2499S}
{Swiderski}, L., {Moszynski}, M., {Nassalski}, A., {Syntfeld-Kazuch}, A., {Szczesniak}, T., {Kamada}, K., {Tsutsumi}, K., {Usuki}, Y., {Yanagida}, T., {Yoshikawa}, A., {Chewpraditkul}, W., and {Pomorski}, M., ``{Scintillation Properties of Praseodymium Doped LuAG Scintillator Compared to Cerium Doped LuAG, LSO and LaBr$_3$},'' {\em IEEE Transactions on Nuclear Science}~{\bf 56},  2499--2505 (Aug. 2009).

\bibitem{1961NucIM..11..340I}
{Iredale}, P., ``{The effect of the non-proportional response of NaI(Tl) crystals to electrons upon the resolution for {\ensuremath{\gamma}}-rays},'' {\em Nuclear Instruments and Methods}~{\bf 11},  340--346 (Jan. 1961).

\bibitem{1995RadM...24..355D}
{Dorenbos}, P., {De Haas}, J.~T.~M., and {Van Eijk}, C.~W.~E., ``{Non-proportional response of scintillation crystals to X-rays and {\ensuremath{\gamma}}-rays},'' {\em Radiation Measurements}~{\bf 24},  355--358 (Oct. 1995).

\bibitem{2003NIMPA.506..250A}
{Agostinelli}, S. et~al., ``{GEANT4{\textemdash}a simulation toolkit},'' {\em Nuclear Instruments and Methods in Physics Research A}~{\bf 506},  250--303 (July 2003).

\bibitem{2006ITNS...53..270A}
{Allison}, J. et~al., ``{Geant4 developments and applications},'' {\em IEEE Transactions on Nuclear Science}~{\bf 53},  270--278 (Feb. 2006).

\bibitem{2016NIMPA.835..186A}
{Allison}, J. et~al., ``{Recent developments in GEANT4},'' {\em Nuclear Instruments and Methods in Physics Research A}~{\bf 835},  186--225 (Nov. 2016).

\bibitem{farrell2008characterising}
Farrell, D.~J., {\em Characterising the performance of luminescent solar concentrators}, PhD thesis, Imperial College London (2008).

\bibitem{2006NIMPA.564..506V}
{Vilardi}, I., {Braem}, A., {Chesi}, E., {Ciocia}, F., {Colonna}, N., {Corsi}, F., {Cusanno}, F., {De Leo}, R., {Dragone}, A., {Garibaldi}, F., {Joram}, C., {Lagamba}, L., {Marrone}, S., {Nappi}, E., {S{\'e}guinot}, J., {Tagliente}, G., {Valentini}, A., {Weilhammer}, P., and {Zaidi}, H., ``{Optimization of the effective light attenuation length of YAP:Ce and LYSO:Ce crystals for a novel geometrical PET concept},'' {\em Nuclear Instruments and Methods in Physics Research A}~{\bf 564},  506--514 (Aug. 2006).

\bibitem{2013NIMPA.712...34I}
{Iwanowska}, J., {Swiderski}, L., {Szczesniak}, T., {Sibczynski}, P., {Moszynski}, M., {Grodzicka}, M., {Kamada}, K., {Tsutsumi}, K., {Usuki}, Y., {Yanagida}, T., and {Yoshikawa}, A., ``{Performance of cerium-doped Gd$_{3}$Al$_{2}$Ga$_{3}$O$_{12}$ (GAGG:Ce) scintillator in gamma-ray spectrometry},'' {\em Nuclear Instruments and Methods in Physics Research A}~{\bf 712},  34--40 (June 2013).

\bibitem{wanarak2012light}
Wanarak, C., Chewpraditkul, W., and Phunpueok, A., ``{Light Yield Non-proportionality and Energy Resolution of Lu$_{1.95}$Y$_{0.05}$SiO$_5$:Ce and Lu$_2$SiO$_5$:Ce Scintillation Crystals},'' {\em Procedia Engineering}~{\bf 32},  765--771 (2012).

\bibitem{1986JaJAP..25.1435I}
{Ishibashi}, H., {Akiyama}, S., and {Ishii}, M., ``{Effect of Surface Roughness and Crystal Shape on Performance of Bismuth Germanate Scintillators},'' {\em Japanese Journal of Applied Physics}~{\bf 25},  1435 (Sept. 1986).

\bibitem{2018PMB....63k5011C}
{Cates}, J.~W. and {Levin}, C.~S., ``{Evaluation of a clinical TOF-PET detector design that achieves {\ensuremath{\leqslant}}100 ps coincidence time resolution},'' {\em Physics in Medicine and Biology}~{\bf 63},  115011 (June 2018).

\end{thebibliography}
\bibliographystyle{spiebib} 

\end{document}